\documentclass[iop]{emulateapj}

\shorttitle{MOIRCS Deep Survey. VIII.}
\shortauthors{Kajisawa et al.}

\begin{document}


\title{MOIRCS Deep Survey. VIII. Evolution of Star Formation Activity as a Function of Stellar Mass in Galaxies since $z\sim3$} 
\author{M. Kajisawa\altaffilmark{1 2},  T. Ichikawa\altaffilmark{2}, T. Yamada\altaffilmark{2}, Y. K. Uchimoto\altaffilmark{3}, T. Yoshikawa\altaffilmark{4}, M. Akiyama\altaffilmark{2}, M. Onodera\altaffilmark{5}}
%T. Ichikawa\altaffilmark{1}, I. Tanaka\altaffilmark{2}, M. Konishi\altaffilmark{3}, T. Yamada\altaffilmark{1}, M. Akiyama\altaffilmark{1},\\ R. Suzuki\altaffilmark{2}, C. Tokoku\altaffilmark{1}, Y. K. Uchimoto\altaffilmark{3}, T. Yoshikawa\altaffilmark{1}, M. Ouchi\altaffilmark{4}, I. Iwata\altaffilmark{5},\\ T. Hamana\altaffilmark{6}, M. Onodera\altaffilmark{7}}
%\affil{Astronomy Department, University of California,
%    Berkeley, CA 94720}
%
%\author{C. D. Biemesderfer\altaffilmark{4,5}}
%\affil{National Optical Astronomy Observatories, Tucson, AZ 85719}
\email{kajisawa@cosmos.ehime-u.ac.jp}
%
%\and
%
%\author{R. J. Hanisch\altaffilmark{5}}
%\affil{Space Telescope Science Institute, Baltimore, MD 21218}


\altaffiltext{1}{Research Center for Space and Cosmic Evolution, Ehime University, Bunkyo-cho 2-5, Matsuyama 790-8577, Japan}
\altaffiltext{2}{Astronomical Institute, Tohoku University, Aramaki,
Aoba, Sendai 980--8578, Japan}
%\altaffiltext{2}{Subaru Telescope, National Astronomical Observatory
%of Japan, 650 North Aohoku Place, Hilo, HI 96720, USA}
\altaffiltext{3}{Institute of Astronomy, University of Tokyo, Mitaka, Tokyo
181--0015, Japan}
\altaffiltext{4}{Koyama Astronomical Observatory, Kyoto Sangyo University, Motoyama, Kamigamo, Kita-ku, Kyoto 603--8555, Japan}
%\altaffiltext{4}{Observatories of the Carnegie Institution of Washington, 813 Santa Barbara Street, Pasadena, CA 91101, USA}
%\altaffiltext{5}{Okayama Astrophysical Observatory, National Astronomical Observatory of Japan, Kamogata, Asakuchi, Okayama, 719--0232, Japan}
%\altaffiltext{6}{National Astronomical Observatory of Japan, Mitaka, Tokyo
%181--8588, Japan}
\altaffiltext{5}{Service d'Astrophysique, CEA Saclay, Orme des Merisiers, 91191 Gif-sur-Yvette Cedex, France}


\begin{abstract}
We study the evolution of star formation activity of galaxies 
at $0.5<z<3.5$ as 
 a function of stellar mass, using very deep NIR data taken with 
Multi-Object Infrared Camera and Spectrograph (MOIRCS) on the Subaru telescope 
in the GOODS-North region. 
The NIR imaging data reach $K \sim$ 23--24 Vega magnitude and 
they allow us to construct a nearly stellar mass-limited 
sample down to $\sim 10^{9.5-10}$ M$_{\odot}$ even at $z\sim3$.
We estimated star formation rates (SFRs) of the sample with two indicators, namely, 
the Spitzer/MIPS 24$\mu$m flux and the rest-frame 2800 \AA\ luminosity.
The SFR distribution at a fixed M$_{\rm star}$ 
shifts to higher values with increasing redshift at $0.5<z<3.5$. 
More massive galaxies show stronger evolution of SFR at $z\gtrsim 1$. 
We found galaxies at $2.5<z<3.5$ show a bimodality in their SSFR distribution, 
which can be divided into two populations by a constant SSFR of 
$\sim$ 2 Gyr$^{-1}$. Galaxies in the low-SSFR group have SSFRs of $\sim $ 
0.5--1.0 Gyr$^{-1}$, while the high-SSFR population shows $\sim$ 10 Gyr$^{-1}$.
The cosmic SFRD is dominated by galaxies 
with M$_{\rm star} = 10^{10-11}$ M$_{\odot}$ at $0.5<z<3.5$, 
while the contribution of 
massive galaxies with M$_{\rm star} = 10^{11-11.5}$ M$_{\odot}$ shows a 
strong evolution at $z>1$ and becomes significant at $z\sim3$, especially 
in the case with the SFR based on MIPS 24 $\mu$m. 
In galaxies with M$_{\rm star} = 10^{10-11.5}$ M$_{\odot}$, 
those with a 
relatively narrow range of SSFR ($\lesssim$ 1 dex) dominates the 
cosmic SFRD at $0.5<z<3.5$. 
The SSFR of galaxies which dominate the SFRD systematically increases 
with redshift. 
At $2.5<z<3.5$, the high-SSFR population,  
which is relatively small in number, dominates the SFRD. 
Major star formation in the universe at higher redshift seems to be 
associated with a more rapid growth of stellar mass of galaxies.
\end{abstract}


\keywords{galaxies: evolution --- galaxies: high-redshift --- infrared:galaxies}



\section{Introduction}
Determining how galaxies build up their stellar mass is crucial for 
understanding galaxy formation and evolution. 
The star formation rate (SFR) and the stellar mass, which is approximately 
considered as the integral of the past SFR, are the most important 
physical properties of galaxies to investigate their stellar mass 
assembly history. Many previous studies measured 
 the comoving cosmic SFR density as a function of redshift, 
which provides a global picture of star formation history in the universe,
using the rest-frame UV luminosity (e.g., \citealp{lil96}; \citealp{mad96}; 
\citealp{con97}; \citealp{mad98}; \citealp{cow99}; \citealp{ste99}; 
\citealp{sch05}; \citealp{bou07}; \citealp{red08}), 
the mid-infrared (MIR) flux (e.g., \citealp{lef05}; \citealp{per05}; 
\citealp{bab06}; \citealp{cap07}; \citealp{mag09}), 
the optical nebular emission lines 
(\citealp{yan99}; \citealp{ly07}; \citealp{gea08}; 
\citealp{shi09}; \citealp{dal10}).
These studies revealed that the cosmic SFR density increases by an order 
of magnitude from $z\sim0$ to $z\sim1$, and then 
reaches a peak around $z\sim2$ and 
decreases at higher redshift (\citealp{hop04}; \citealp{hop06}).
On the other hand, recent near-infrared (NIR) surveys enable to 
measure the average stellar mass density of the universe and several studies 
based on deep NIR surveys found that a rapid evolution of the cosmic 
stellar mass density occurred at $1 \lesssim z \lesssim 3$ (e.g., 
\citealp{dic03}; \citealp{fon06}; \citealp{kaj09} and reference therein).
These results suggest that a significant fraction of the stellar mass in 
the present universe had been formed at $z\sim$ 1--3.

The next step to understand how star formation and stellar mass assembly 
proceeded in galaxies at $1 \lesssim z \lesssim 3$ is to investigate the 
 SFR of these galaxies 
as a function of stellar mass and redshift.
Multi-wavelength data from surveys such as COMBO-17, GOODS, COSMOS, UKIDSS, 
and AEGIS, have been used to measure SFR and stellar mass of galaxies 
at $1 \lesssim z \lesssim 3$
 in many studies, where SFRs are estimated from various indicators,
namely, the rest-frame UV luminosity (\citealp{jun05}; 
\citealp{dad07}; \citealp{feu07}; 
\citealp{cow08}; \citealp{dro08}; \citealp{mob09}; \citealp{mag10}), 
the Spitzer/MIPS 24 $\mu$m flux (\citealp{pap06}; 
\citealp{elb07}; \citealp{noe07a};  
\citealp{zhe07}; \citealp{san09}; \citealp{dam09a}; \citealp{dam09b}), 
the H$\alpha$ emission line (\citealp{erb06}; \citealp{for09}), 
and the radio continuum emission (\citealp{pan09}; \citealp{dun09}).
These studies found that 
the average SFR or SSFR of 
star-forming galaxies at a fixed stellar mass 
increases with redshift strongly. 
The rate of incline is similar in all stellar mass ranges investigated 
(\citealp{zhe07}; \citealp{dam09a}; \citealp{dam09b}).
The SSFR of star-forming galaxies at $z\sim2$ is higher 
than that at $z\sim1$ by factor of several and that at $z\sim0$ by 
factor of several tens (\citealp{dad07}; \citealp{san09}; \citealp{pan09}).
In the studies at $z\sim2$, however, the sample selection is based on 
relatively shallow NIR/MIR data or deep but optical data, and therefore 
the sample becomes incomplete at relatively high stellar mass. 
For example, a depth of 23--24 AB magnitude in the NIR--MIR wavelength 
corresponds to 
a limiting stellar mass of $\sim 10^{10-10.5}$ M$_{\odot}$ 
at $z\sim2$ (e.g., \citealp{dun09}). 
On the other hand, the selection in the optical band 
could preferentially pick up actively star-forming galaxies and could miss objects 
with relatively low SFR at low mass (e.g., \citealp{feu07}; \citealp{kaj06b}). 
If this is the case, the average SFR at low mass would be overestimated. 
In order to investigate the mass dependence of star formation activity of 
galaxies at $z\sim2$ precisely, the selection based on more deep NIR/MIR data is 
desirable.

On the other hand, the SFR as a function of stellar mass and redshift 
enables to estimate 
the contribution of galaxies in different stellar masses to the cosmic 
SFR density and its evolution. 
At $z \lesssim 2$, galaxies with stellar mass of $\sim 10^{10-11}$ M$_{\odot}$ 
dominate the cosmic SFR density, and the decrease of 
the contribution of these galaxies from $z\sim2$ to the present seems to 
result in the evolution of the total SFR density of the universe 
(\citealp{cow08}; \citealp{mob09}; \citealp{san09}). 
The evolution of the 
contribution to the cosmic SFR density is also similar in all mass 
ranges at $z\lesssim 1$. Several studies suggest that
 the contribution of massive galaxies 
with M$_{\rm star} > 10^{11}$ M$_{\odot}$ increases with redshift 
 more strongly than lower-mass 
galaxies at $z>1$, although these massive galaxies do not 
dominate the total SFR density even at $z\sim2$ (\citealp{jun05}; \citealp{san09}).
As mentioned above, however, only galaxies with relatively high stellar mass tend to 
be sampled in these studies,  and therefore 
the evolution of the contribution of relatively 
low-mass galaxies is still unclear.  
Furthermore, the similar analysis for galaxies at $z\sim3$ has not been performed 
due to the lack of sufficiently deep NIR/MIR data.
Since the cosmic SFRD at $z\sim3$ is estimated to be similarly high 
 with that at $z\sim$ 1--2 \citep{hop06},  
it is important to study how star formation occurs 
in galaxies with different stellar masses at $z\sim3$.

In this paper, we study the 
SFR as a function of stellar mass for galaxies at 
$0.5<z<3.5$ in the GOODS-North field, using very deep NIR data  
 from MOIRCS Deep Survey (MODS, \citealp{kaj06a}; \citealp{ich07}). 
The MODS data reach $\sim$ 23--24 Vega magnitude ($\sim$ 25--26 AB magnitude) 
in the $K$ band, and they allow us to construct a stellar mass 
limited sample down to $\sim 10^{9.5-10}$ M$_{\odot}$ even at $z\sim3$ 
and to investigate the distribution of the SFR at high redshift 
without biases for objects with very high SFR (approximately 
low stellar mass-to-light ratio) at low mass. 
We use the MODS data and publicly available multi-wavelength data 
of the GOODS survey to 
estimate redshift, SFR, and stellar mass of the sample.
Section 2 describes the data set and the procedures of source detection and 
photometry. In Section 3, we construct the stellar mass limited sample.
The methods for estimating the SFRs of the sample are presented in Section 4.
We show the distribution of the SFR as a function of stellar mass and 
its evolution in Section 5, and  
discuss their implication in Section 6. 
A summary is provided in Section 7.

We use a cosmology with H$_{\rm 0}$=70 km s$^{-1}$ Mpc$^{-1}$, 
$\Omega_{\rm m}=0.3$ and $\Omega_{\rm \Lambda}=0.7$.
The Vega-referred magnitude system is used throughout this paper, 
unless stated otherwise.

%%%%%%%%%%%%%%%%%%%%%%%%%
\begin{table*}
\begin{center}
%% \begin{minipage}{140mm}
  \caption{Number of galaxies in each redshift bin}
  \label{tab:num}
  \begin{tabular}{@{}lccccc   @{}}
  \tableline 
  \tableline
& redshift &  Wide\tablenotemark{a} & Deep\tablenotemark{a} & combined\tablenotemark{a} & combined \\
& & (103arcmin$^{2}$ $K<23$)&(28.2arcmin$^2$ $K<24$)& & (MIPS 24$\mu$m-detected) \\  
 \tableline
& 0.5--1.0 &  1618 (843) & 702 (306) & 1701 (843) & 626 \\
& 1.0--1.5 & 1156 (348) & 481 (105) & 1241 (348) & 408 \\
& 1.5--2.5 & 1008 (189) & 469 (75) & 1109 (189) & 417 \\
& 2.5--3.5 & 307 (65) & 234 (47) & 370 (66) & 84 \\

\tableline
\label{table:sample}
\end{tabular}
\tablenotetext{a}{Number in the parenthesis indicates objects with spectroscopic redshift.}
%%\end{minipage}
\end{center}
\end{table*}
%%%%%%%%%%%%%%%%%%%%%%%%%%%%%%%%%%%
\section{Observational Data and Photometry}
\label{sec:obs}
We use the $K$-selected sample of the MODS in the GOODS-North 
region (\citealp{kaj09}, hereafter K09), 
which is based on our deep $JHK_{s}$-bands 
imaging data taken with MOIRCS \citep{suz08} on the Subaru telescope.
Four MOIRCS pointings cover $\sim70$\% of the GOODS-North region ($\sim$ 103.3
arcmin$^2$, hereafter referred as ``wide'' field) 
and the data reach $J=24.2$, $H=23.1$, $K=23.1$ (5$\sigma$, 
Vega magnitude). One of the four pointings 
is the ultra-deep field of the MODS ($\sim$ 28.2 arcmin$^2$, hereafter ``deep''
 field),  
where the data reach $J=25.1$, $H=23.7$, $K=24.1$. A full description of the 
observations, reduction, and quality of the data is presented in a 
separate paper \citep{kaj10}. 

The source detection was performed in the $K$-band image using the SExtractor
image analysis package \citep{ber96}. At first, we limited the samples 
to $K<23$ and $K<24$ for the wide and deep fields, where the detection 
completeness for point sources is more than 90\%. Then we measured 
the optical-to-MIR SEDs of the sample objects, using the publicly available
multi-wavelength data in the GOODS field, namely KPNO/MOSAIC ($U$ band, 
\citealp{cap04}), $Hubble Space Telescope$/Advanced Camera for Surveys 
($HST$/ACS; $B$, $V$, $i$, $z$ bands, version 2.0 data; M. Giavalisco et al. 
2010, in preparation; \citealp{gia04}) and $Spitzer$/IRAC (3.6$\mu$m, 4.5$\mu$m,
 5.8$\mu$m, DR1 and DR2; M. Dickinson et al. 2010, in preparation), as well as 
the MOIRCS $J$- and $H$-bands images. Details of the multi-band aperture 
photometry are presented in K09. Following K09, 
we used objects which are detected above 
2$\sigma$ level in more than two other bands in addition to the 5$\sigma$ detection
in the $K$-band,  
because it is difficult to estimate the photometric redshift and stellar mass of 
those detected only in one or two bands. The number of those excluded by 
this criterion is negligible (21/6402 and 42/3203 for the wide and deep fields, 
respectively).

%% In a manner similar to \objectname authors can provide links to dataset
%% hosted at participating data centers via the \dataset{} command.  The
%% second curly bracket argument is printed in the text while the first
%% parentheses argument serves as the valid data set identifier.  Large
%% lists of data set are best provided in a table (see Table 3 for an example).
%% Valid data set identifiers should be obtained from the data center that
%% is currently hosting the data.
%%
%% Note that AASTeX interprets everything between the curly braces in the 
%% macro as regular text, so any special characters, e.g. "#" or "_," must be 
%% preceded by a backslash. Otherwise, you will get a LaTeX error when you 
%% compile your manuscript.  Special characters do not 
%% need to be escaped in the optional, square-bracket argument.

%% In this section, we use  the \subsection command to set off
%% a subsection.  \footnote is used to insert a footnote to the text.


%% This section also includes several of the displayed math environments
%% mentioned in the Author Guide.

\section{Sample Selection}
\label{sec:sample}
K09 estimated the redshift and stellar mass of 
the $K$-selected galaxies mentioned above and 
constructed a stellar mass-limited sample to study the evolution of 
the stellar mass function. We use the same stellar mass-limited sample in 
this study and describe briefly how the sample was constructed in the 
following.

In order to estimate the photometric redshift and stellar mass, K09 
performed spectral energy distribution (SED) 
fitting of the multi-band photometry
described above ($UBVizJHK$, 3.6$\mu$m, 4.5$\mu$m, and 5.8$\mu$m) with 
population synthesis models. We here adopt the results with GALAXEV model 
\citep{bru03}. We used the model templates with exponentially decaying 
star formation histories with the decaying timescale $\tau$ ranging between 
0.1 and 20 Gyr and Calzetti extinction law \citep{cal00} in the range of 
$E(B-V)=$ 0.0-1.0. Metallicity is changed from 1/50 to 1 solar metallicity.
We assume Salpeter IMF \citep{sal55} with lower and upper 
mass limits of 0.1 and 100 M$_{\odot}$ for easy comparison with the results
in other studies.
If we assume the Chabrier-like IMF \citep{cha03}, the stellar mass is reduced 
by a factor of $\sim$ 1.8. 
The model age is changed from 50 Myr 
to the age of the universe at the observed redshifts.
The resulting photometric redshifts show a good agreement with spectroscopic 
redshifts (Figure 1 in K09). If available, we adopted spectroscopic redshifts
from the literature (\citealp{coh00}; \citealp{coh01}; \citealp{daw01}; 
\citealp{wir04}; \citealp{cow04}; \citealp{tre05}; \citealp{cha05}; 
\citealp{red06}; \citealp{bar08}; \citealp{yos10}), and performed the 
SED fitting fixing the redshift to each spectroscopic value for these galaxies.
The estimated redshifts and stellar mass-to-luminosity (M/L) ratios from 
the best-fit templates were used to calculate the stellar mass.
The uncertainty of the stellar mass is estimated taking into account of 
the photometric redshift error for those 
without spectroscopic redshift, and is discussed 
in detail in K09.

K09 also investigated how the $K$-band magnitude limit affects the stellar 
mass distribution of the sample by 
using the distribution of the rest-frame $U-V$ color, which reflects 
the stellar M/L ratio well, as a function of stellar mass, and then 
determined the limiting stellar mass above which more than 90\% of 
objects are expected to be brighter than the $K$-band magnitude limit 
 as a function of redshift (Figure 3 and 4 in K09). 
The limiting stellar mass is $\sim 1\times 10^{9}$ M$_{\odot}$ ($\sim 3\times 10^{8}$ M$_{\odot}$) for the wide (deep) sample at $z=0.75$, $\sim 2\times 10^{9}$ 
M$_{\odot}$ ($\sim 8\times 10^{8}$ M$_{\odot}$) at $z=1.25$, 
$\sim 5\times 10^{9}$ M$_{\odot}$ ($\sim 2\times 10^{9}$ M$_{\odot}$) at $z=2.0$,
$\sim 1\times 10^{10}$ M$_{\odot}$ ($\sim 3\times 10^{9}$ M$_{\odot}$) at $z=3.0$,
 respectively.
We use only objects above this limiting mass in 
this study. Further details of determining the limiting stellar mass are 
given in K09. 

Table \ref{tab:num}
 lists the number of objects in each redshift bin. We use the same 
redshift bins as those used in K09. The bin width is sufficiently 
larger than the 
typical photometric errors, and each redshift bin includes a reasonable 
number of galaxies for studying the SFR as a function of 
stellar mass statistically. In the following analysis, 
we basically use ``combined'' (wide $+$ 
deep) sample except for the estimates of the comoving number density and 
SFR density (section \ref{sec:ssfrms} and \ref{sec:sfrd}), 
where we need to calculate V$_{\rm max}$ separately for 
the wide and deep samples.

%% The equation environment wil produce a numbered display equation.

%% The \notetoeditor{TEXT} command allows the author to communicate
%% information to the copy editor.  This information will appear as a
%% footnote on the printed copy for the manuscript style file.  Nothing will
%% appear on the printed copy if the preprint or
%% preprint2 style files are used.

%% The eqnarray environment produces multi-line display math. The end of
%% each line is marked with a \\. Lines will be numbered unless the \\
%% is preceded by a \nonumber command.
%% Alignment points are marked by ampersands (&). There should be two
%% ampersands (&) per line.

%% Putting eqnarrays or equations inside the mathletters environment groups
%% the enclosed equations by letter. For instance, the eqnarray below, instead
%% of being numbered, say, (4) and (5), would be numbered (4a) and (4b).
%% LaTeX the paper and look at the output to see the results.

%% This section contains more display math examples, including unnumbered
%% equations (displaymath environment). The last paragraph includes some
%% examples of in-line math featuring a couple of the AASTeX symbol macros.

\section{Estimate of the SFR}
\label{sec:sfr} 
In order to estimate the SFRs associated with the sample galaxies, 
we use two SFR indicators, namely, rest-frame 2800 \AA\ luminosity and observed 
24$\mu$m flux. The rest-frame UV luminosity probes the light from young 
massive stars, while the IR luminosity measures those photons 
absorbed and re-emitted by dust. We basically use a combination of the IR and 
(dust-uncorrected) UV luminosities for objects with Spitzer/MIPS 24$\mu$m 
detection and adopt dust-corrected UV luminosity for those without the MIPS 
detection. For comparison, we also check the case where the dust-corrected 
UV luminosity is used for all galaxies.

%--------------------------figures
\begin{figure*}
\begin{center}
\includegraphics[angle=0,scale=1.0]{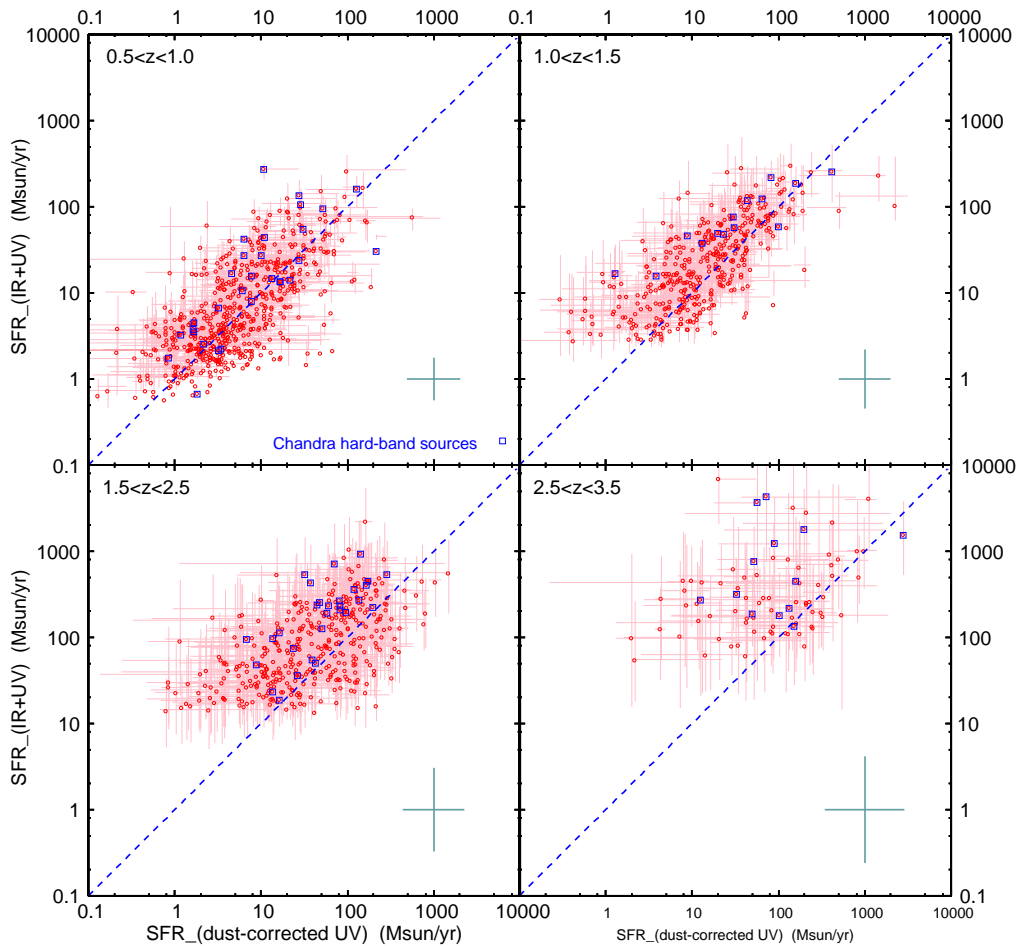}
\end{center}
\caption{Comparison between SFR$_{\rm IR+UV}$ and SFR$_{\rm corrected-UV}$ for 
galaxies detected in the MIPS 24$\mu$m image for each redshift bin. 
Squares represent X-ray sources detected in the Chandra hard-band image.
Typical errors of 
the SFR$_{\rm IR+UV}$ and SFR$_{\rm corrected-UV}$ are shown at 
the bottom right corner of each panel.
}
\label{fig:compsfr}
\end{figure*}
%%%%%%%%%%%%%%%%%%%%%%%%%%%%%%%%%%%%
\subsection{SFR based on MIPS 24$\mu$m flux} 
\label{sec:irsfr} 
Public MIPS 24$\mu$m data for 
the GOODS-North (DR1+, M. Dickinson et al., in preparation) was used for 
the MIR photometry. Since this image is very deep and its PSF size is 
relatively large (FWHM $\sim $ 5.4 arcsec), the source confusion occurs for 
many objects. Following \cite{lef05},   
we use the IRAF/DAOPHOT package \citep{ste87} 
in order to deal with these blended sources properly.
The DAOPHOT software is based on the PSF fitting technique and allows us to  
fit blended sources in a crowded region simultaneously and then derive 
the flux of each object from the scaled fitted PSF. An empirical PSF 
was constructed from bright isolated point sources. We used the positions 
of the sample galaxies on the high-resolution MOIRCS $K_{s}$-band 
image as a prior
for the centers of the fitted PSFs in the photometry. 
The background noise was estimated by measuring sky fluxes at 
 random positions on the image. For sources whose residual of the PSF fitting 
was significantly larger than the background noise, we added
 the fitting residual to the photometric error. 
We set a detection threshold of 5$\sigma$, which corresponds to 
$\sim 20\mu$Jy in most cases where the residual of the fitting is negligible.

In order to convert the MIPS 24$\mu$m flux to the total IR luminosity 
(8-1000 $\mu$m), we used SED templates of star-forming galaxies 
provided by \cite{dal02}. The model templates cover a wide range of SED shapes,
allowing for different heating levels of the interstellar dust by the radiation
field. Following \cite{wuy08} and \cite{dam09a}, we calculated the total IR 
luminosity for each object using the all templates with the 
range from $\alpha=1$ to $\alpha=2.5$. $\alpha$ is defined by 
$dM(U)\sim U^{-\alpha}dU$, where $M(U)$ represents the dust mass headed by an 
intensity U of the radiation field \citep{dal02}. 
Then we averaged the calculated values for the different $\alpha$ in a
 logarithmic scale and adopted as a best estimate for the IR luminosity.
The variation from the value for $\alpha=1$ to that for $\alpha=2.5$ was taken  
into account in the computation of the error of the total IR luminosity. 
For objects without spectroscopic redshift, we also estimated the effect of 
the uncertainty of the photometric redshift by calculating the IR luminosity 
over the possible ranges of redshift, and added it to the error.

We converted the derived total IR luminosity to a SFR$_{\rm IR}$ using the 
\cite{ken98}'s relation 
\begin{equation}
SFR_{\rm IR}\ (M_{\odot}\ yr^{-1}) = 4.5\times10^{-44}\ L_{\rm IR} (erg\ s^{-1}) .
\end{equation}
Salpeter IMF with lower and upper 
mass limits of 0.1 and 100 M$_{\odot}$ is assumed.
The limiting flux of $\sim 20 \mu$Jy mentioned above corresponds to $\sim 7 
$ M$_{\odot}$/yr at $z=1$, $\sim 40$ M$_{\odot}$/yr at $z=2$, 
and $\sim 300$ M$_{\odot}$/yr at $z=3$. The number of objects 
detected in the MIPS 24 $\mu$m 
image in each redshift bin is 626 at $0.5<z<1.0$, 
408 at $1.0<z<1.5$, 417 at $1.5<z<2.5$, and 
84 at $2.5<z<3.5$, respectively.

\subsection{SFR based on rest-frame 2800 \AA \  luminosity}
\label{sec:uvsfr}
The rest-frame 2800 \AA\ luminosity for each object was calculated from 
the same best-fit SED template as used in the estimate of the photometric
redshift and stellar mass. Since our multi-band photometric data 
cover completely the rest-frame 2800 \AA\ for $0.5<z<3.5$, 
the estimated rest-frame 2800 \AA\ luminosity is a robust quantity, 
especially in the case where a galaxy is detected in both bands which straddle
the rest-frame 2800 \AA. We calculated the confidence interval for it 
in the SED fitting procedure. 
In order to derive 
the dust-corrected UV luminosity, we also used E(B-V) of the 
same best-fit SED template. 
Since E(B-V) is often degenerated with other parameters of stellar population 
such as age and metallicity in the SED fitting, 
the dust-corrected UV luminosity tends to have  
larger errors than the dust-uncorrected one.
Note that our lower limit of the model age of 50 Myr in the SED fitting
affects the estimated E(B-V), thus the dust-corrected UV luminosity and SFR.
We will discuss the effect of the age limit on SFR in Section \ref{sec:sfrms}. 
%--------------------------figures
\begin{figure*}
\begin{center}
\includegraphics[angle=0,scale=1.0]{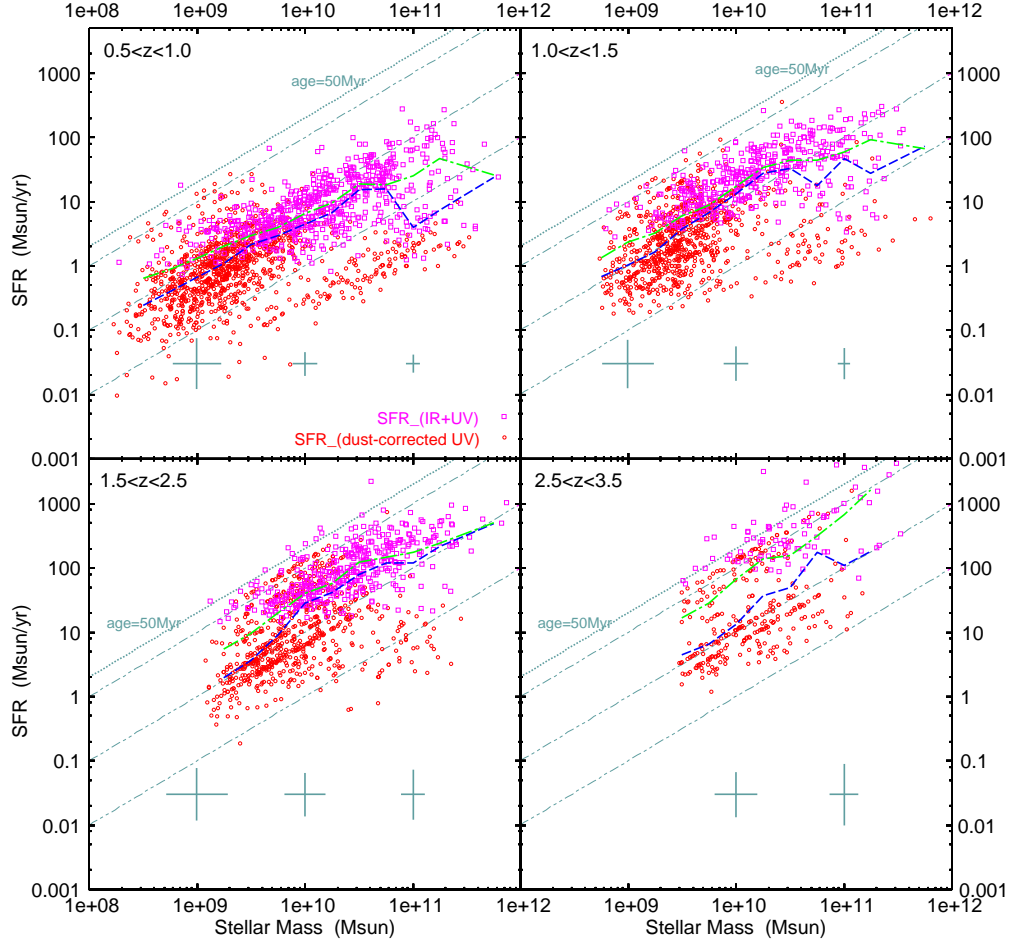}
\end{center}
\caption{SFR vs. M$_{\rm star}$ for our stellar mass limited sample for 
each redshift bin. Squares show galaxies with SFR$_{\rm IR+UV}$ derived from 
the 24 $\mu$m fluxes, while circles represent objects undetected in the 
MIPS 24 $\mu$m image, with SFR$_{\rm corrected-UV}$ derived from the 
dust-corrected rest-frame UV luminosity. 
Median errors of the SFR and stellar mass at each mass are shown 
as the crosses at the bottom of each panel. 
Dotted lines represent SFR/M$_{\rm star} = $ 1/50 Myr$^{-1}$ (see text for 
details).
Double-dotted dashed lines show SFR/M$_{\rm star} = $ 0.1, 1, 10 Gyr$^{-1}$.
Blue dashed lines show the median SFR at each stellar mass.
Green dotted-dashed lines represent the results of the linear fitting for 
star-forming galaxies. In the panel of the $2.5<z<3.5$ bin, the lower 
line shows the result for galaxies with $\log$(SFR/M$_{\rm star}$) $>$ -0.5, and 
upper one shows that for galaxies with $\log$(SFR/M$_{\rm star}$) $>$ 0.25.
}
\label{fig:sfrmsir}
\end{figure*}
%%%%%%%%%%%%%%%%%%%%%%%%%%%%%%%%%%%%

We converted the dust-uncorrected and dust-corrected rest-frame 2800 \AA\ 
luminosities into SFRs using the calibration by \cite{ken98} 
\begin{equation}
SFR_{\rm UV}\ (M_{\odot}\ yr^{-1}) = 1.4\times10^{-28}\ L_{\nu} (erg\ s^{-1}\ Hz^{-1}) .
\end{equation}
The SFR estimated from the dust-uncorrected UV luminosity 
accounts for the unobscured light from young stars and is added to
the SFR$_{\rm IR}$ estimated from the 24$\mu$m flux for objects detected in  
the MIPS 24$\mu$m image.
We refer to the combined value as SFR$_{\rm IR+UV}$ in the following.   
The SFR estimated from the dust-corrected UV luminosity   
(hereafter SFR$_{\rm corrected-UV}$) can be calculated for all the sample.

\subsection{Comparison between SFR$_{\rm IR+UV}$ and SFR$_{\rm corrected-UV}$}
\label{sec:compsfr}
Figure \ref{fig:compsfr} shows a comparison 
between SFR$_{\rm IR+UV}$ and SFR$_{\rm corrected-UV}$ for galaxies detected 
in the MIPS 24$\mu$m image. 
Given the uncertainty, these two estimates agree well with 
each other especially in low-redshift bins.
At low SFR, however, SFR$_{\rm IR+UV}$ tends to be 
 higher than SFR$_{\rm corrected-UV}$, and 
the systematic offset seems to become larger with redshift.
The mean and standard deviation of 
$\log(SFR_{\rm IR+UV}/SFR_{\rm corrected-UV})$ for all objects with the 
MIPS 24$\mu$m detection are $0.19 \pm 0.39$ at $0.5<z<1.0$, 
$0.25 \pm 0.37$ at $1.0<z<1.5$, $0.37 \pm 0.49$ at $1.5<z<2.5$, and 
$0.63 \pm 0.64$ at $2.5<z<3.5$. 
\cite{fra08} reported a similar trend in the comparison between the 
SFRs derived from the MIPS 24$\mu$m flux and the SED fitting.

In the conversion from the 24$\mu$m flux into 
 the total IR luminosity, 
 we don't take into account the luminosity-dependent shapes of 
the IR SED found in local galaxies (e.g., \citealp{cha01}), 
  following \cite{wuy08} and \cite{dam09a}.
Therefore SFR$_{\rm IR+UV}$ 
could be overestimated for galaxies with low IR luminosity 
(SFR) and underestimated for those with high IR luminosity, if 
the luminosity-dust temperature (IR color) relation 
for nearby galaxies holds for galaxies at higher redshift. 
On the other hand, there are several studies that suggest 
 that high-z active star-forming galaxies do not form 
such a relation and galaxies with L$_{\rm IR}>10^{12}$L$_{\odot}$ (ULIRGs)
at high redshift have cooler average dust temperatures than those in 
the local universe (e.g., \citealp{sym09}; \citealp{muz10}).
Furthermore, it is reported that 
bright 24 $\mu$m sources at $z\sim2$ have higher polycyclic 
aromatic hydrocarbon (PAH) luminosity relative to the total IR luminosity
than the local ULIRGs with similar IR luminosity 
(\citealp{hua09}; \citealp{mur09}). 
Since the MIPS 24 $\mu$m band samples the PAH emissions in 6-9 $\mu$m region 
at $z\gtrsim 1.5$, higher L$_{\rm PAH}$/L$_{\rm IR}$ ratios than those for  
the local ULIRGs could lead to a overestimation of SFR for these high-z 
galaxies 
if one converts 24 $\mu$m fluxes into the SFRs using the 
local luminosity-dust temperature relation.
%--------------------------figures
\begin{figure*}
\begin{center}
\includegraphics[angle=0,scale=1.0]{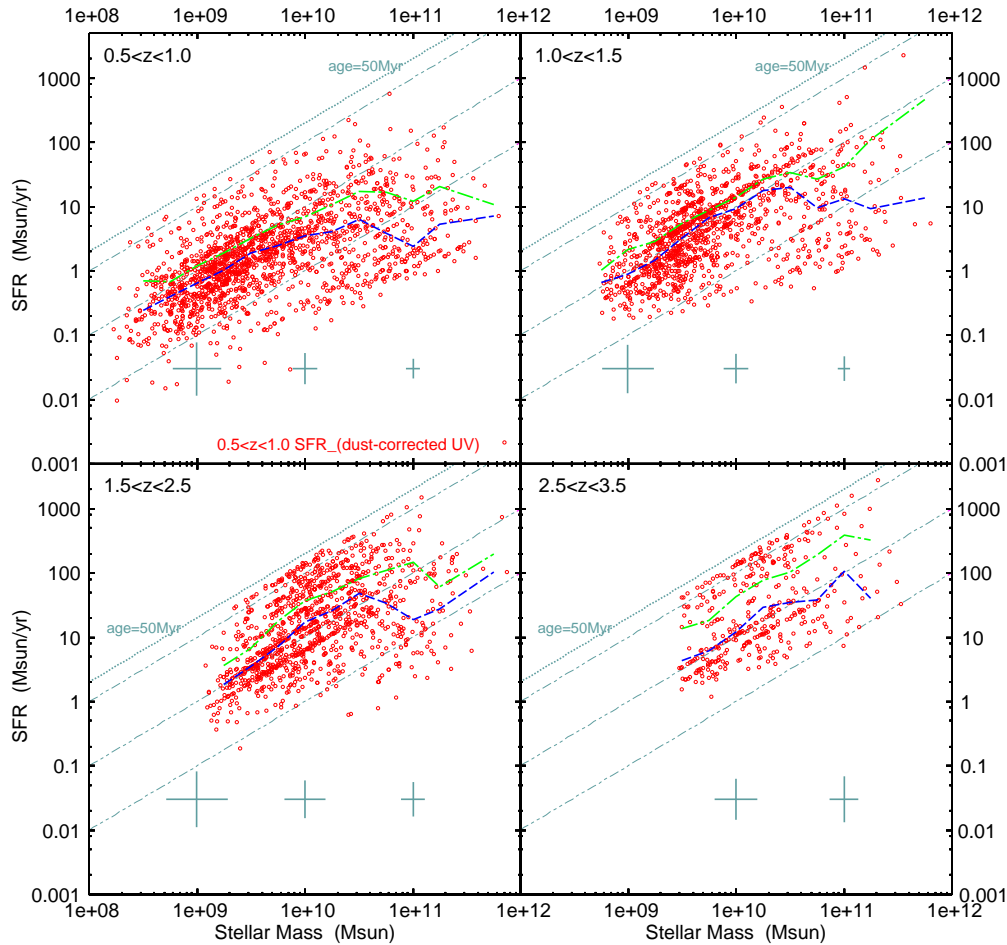}
\end{center}
\caption{The same as Figure \ref{fig:sfrmsir} except that 
SFR$_{\rm corrected-UV}$ is used for all the sample.
}
\label{fig:sfrmsuv}
\end{figure*}
%%%%%%%%%%%%%%%%%%%%%%%%%%%%%%%%%%%%

It should be noted that 
the MIR flux may also be powered partly by AGN (e.g., \citealp{nar08}). 
We showed objects detected in the Chandra hard-band image of 
the CDF-North \citep{ale03} as squares in Figure \ref{fig:compsfr}.
Those objects detected in the hard-band of Chandra 
have slightly higher SFR$_{\rm IR+UV}$/SFR$_{\rm corrected-UV}$ ratios;
the mean and standard deviation of 
$\log(SFR_{\rm IR+UV}/SFR_{\rm corrected-UV})$ for those objects are 
$0.31 \pm 0.38$ at $0.5<z<1.0$,    
$0.37 \pm 0.35$ at $1.0<z<1.5$,    
$0.54 \pm 0.32$ at $1.5<z<2.5$, and   
$0.80 \pm 0.66$ at $2.5<z<3.5$.
SFR$_{\rm IR+UV}$ for those sources could be overestimated due to the 
AGN contribution, although SFR$_{\rm IR+UV}$ tends to be  
higher than SFR$_{\rm corrected-UV}$ especially at high redshift 
even if we exclude the Chandra hard-band sources.

On the other hand, the dust-corrected UV luminosity might be underestimated 
for heavily obscured galaxies such as ULIRGs at low and high redshifts
(e.g., \citealp{gol02}; \citealp{pap06}; \citealp{red10}).
If a galaxy has star-forming regions from which one can detect no UV light at 
all due to the heavy dust obscuration, only UV light from relatively 
less-obscured regions contributes the observed SED, which could result in 
the underestimate of the dust extinction. 

Since the calibration by \cite{ken98} we used 
for the conversion from the rest-frame 2800 \AA\ luminosity into SFR 
 is derived assuming that SFR has remained constant over a timescale of 
$\sim 100$ Myr and that the luminosity is dominated by young O and B stars,  
SFR could be overestimated for relatively quiescent galaxies. 
 For such galaxies, older stellar population than O and B stars could 
contribute significantly to the rest-frame 2800 \AA\ luminosity (e.g., 
\citealp{dal07}). Therefore SFR$_{\rm corrected-UV}$ could be systematically 
overestimated for galaxies with low SFRs relative to their stellar mass.

Keeping in mind the possible systematic errors, 
we show both results with SFR$_{\rm IR+UV}$ 
for objects detected in the  MIPS 24$\mu$m image and 
SFR$_{\rm corrected-UV}$ for 
those without the MIPS detection, and with SFR$_{\rm corrected-UV}$ for 
all the sample in the following section.

%--------------------------figures
\begin{figure*}
\begin{center}
\includegraphics[angle=0,scale=0.9]{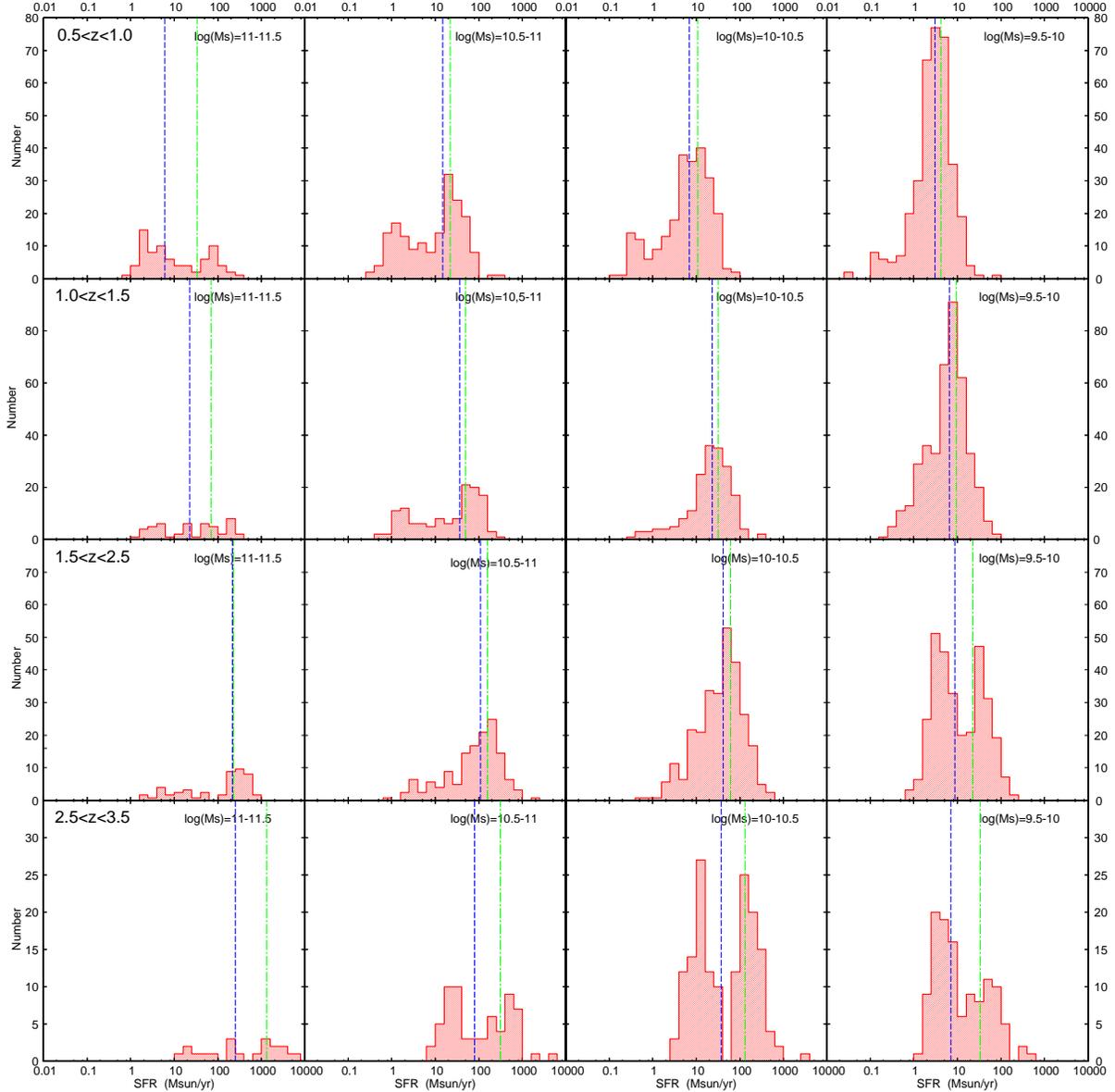}
\end{center}
\caption{SFR distribution for galaxies in each stellar mass and redshift bin.
SFR$_{\rm IR+UV}$ is used if a galaxy is detected in the MIPS 24$\mu$m
 image and SFR$_{\rm corrected-UV}$ is used 
for objects without the 24$\mu$m detection.
Dashed and dotted-dashed lines show the median and average SFRs for 
each mass and redshift bin.
}
\label{fig:histsfrir}
\end{figure*}
%%%%%%%%%%%%%%%%%%%%%%%%%
\section{Results}
\label{sec:result}

\subsection{SFR as a function of Stellar Mass at $0.5<z<3.5$} 
\label{sec:sfrms}
Figure \ref{fig:sfrmsir} and \ref{fig:sfrmsuv} show 
the SFR as a function of M$_{\rm star}$ for the stellar mass selected sample 
 in the four redshift bins.
Figure \ref{fig:sfrmsir} represents the case 
where SFR$_{\rm IR+UV}$ is used if a galaxy is detected in the MIPS 24$\mu$m
 image and SFR$_{\rm corrected-UV}$ is used 
for those without the 24$\mu$m detection,  
and Figure \ref{fig:sfrmsuv} represents 
the case where SFR$_{\rm corrected-UV}$ is used for all the sample. 
A dotted line in each redshift panel represents SFR/M$_{\rm star} = $ 
1/50 Myr$^{-1}$. 
In the calculation of the SFR$_{\rm corrected-UV}$, 
both the rest-frame 2800 \AA\ luminosity and the E(B-V) were estimated from 
the SED fitting. The stellar mass is also based on the same best-fit SED model.
Since SFR/M$_{\rm star}$ of the model SED  
cannot exceed the inverse of the age for the monotonically
decaying SFR models (SFR/M$_{\rm star}$ roughly equals the inverse of the age 
for the constant SFR models if one ignores the stellar mass loss due to the 
super nova or stellar wind), 
the lower limit of the model age leads to the upper limit for the 
SFR/M$_{\rm star}$. Therefore, 
the lower limit of age $>$ 50 Myr in the SED fitting mentioned above causes  
the upper limit of SFR$_{\rm corrected-UV}$/M$_{\rm star} < $ 1/50 Myr$^{-1}$ 
($=$ 20 Gyr$^{-1}$). 
On the other hand, the SFR$_{\rm IR+UV}$ is not affected by this limit, 
because the SFR based on the 24 $\mu$m flux are independent of the result 
of the SED fitting for the stellar emission.  
Since there are only a few objects whose SFR$_{\rm IR+UV}$ significantly 
exceeds this limit in high-redshift bins in Figure \ref{fig:sfrmsir}, 
we believe that the lower limit of the model age does not significantly bias 
the distribution of the SFR$_{\rm corrected-UV}$ as a function of M$_{\rm star}$.

In Figure \ref{fig:sfrmsir} and \ref{fig:sfrmsuv}, 
we plot the median and average of SFR at each mass as a 
dashed and dotted-dashed lines, respectively, 
to examine in detail the stellar mass dependence of the SFR.
At M$_{\rm star} \lesssim 10^{10.5}$ M$_{\odot}$, 
more massive galaxies tend to have higher SFRs in all redshift bins, 
although there is a substantial scatter at each mass. 
The median and average SFRs of these galaxies increase with M$_{\rm star}$. 
On the other hand, 
the trend becomes less clear at 
M$_{\rm star} \gtrsim 10^{10.5}$ M$_{\odot}$;the median SFR does not significantly
increase with M$_{\rm star}$, 
especially in the case with SFR$_{\rm corrected-UV}$ 
for all the sample (Figure \ref{fig:sfrmsuv}). 
There are massive 
galaxies with relatively low SFRs similar with low-mass galaxies, 
while galaxies with higher SFRs which follow the trend at lower M$_{\rm star}$ also 
exist.

%--------------------------figures
\begin{figure*}
\begin{center}
\includegraphics[angle=0,scale=0.9]{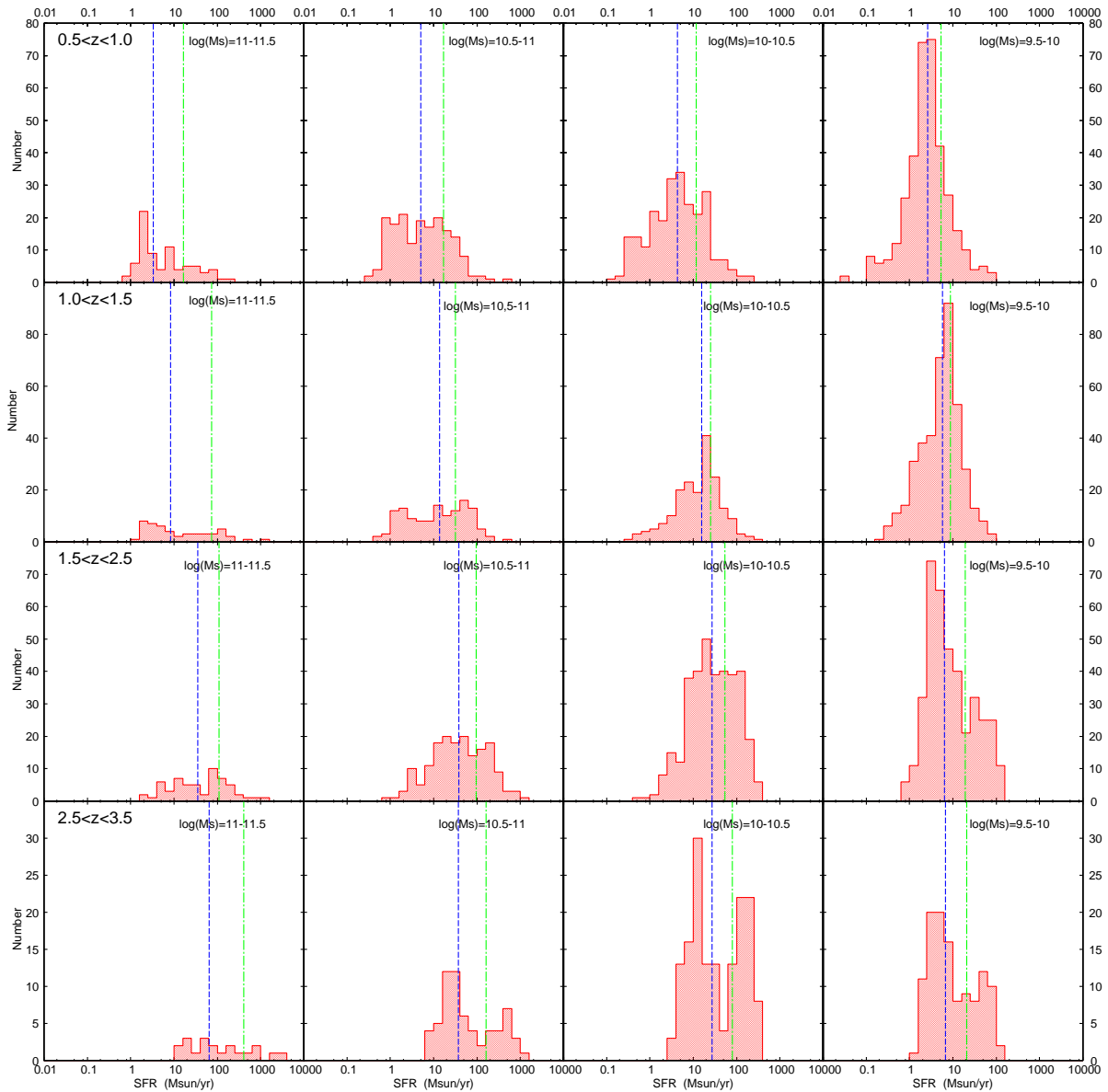}
\end{center}
\caption{
The same as Figure \ref{fig:histsfrir} except that 
SFR$_{\rm corrected-UV}$ is used for all the sample.
}
\label{fig:histsfruv}
\end{figure*}
%%%%%%%%%%%%%%%%%%%%%%%%%
We note that the SFR distribution at a fixed M$_{\rm star}$ 
shifts to higher values with increasing redshift at least at $z<2.5$ 
in Figure \ref{fig:sfrmsir} and \ref{fig:sfrmsuv}.
Figure \ref{fig:histsfrir} and \ref{fig:histsfruv} 
show the SFR distribution for each stellar mass and redshift bin.
At $z<2.5$, the shift of the SFR distribution with redshift is seen in all mass bins in 
the both figures, although there are galaxies with relatively wide range of SFR in 
each mass and redshift bin. Furthermore, 
it is seen that more massive galaxies show stronger evolution of the SFR in 
the both cases with different SFR indicators.

At $2.5<z<3.5$, galaxies with $M_{\rm star} = $
10$^{9.5}$--10$^{11}$ M$_{\odot}$ show a bimodality in the SFR, and the SFR 
distribution does not concentrate around the median or average value 
(Figure \ref{fig:histsfrir} and \ref{fig:histsfruv}). 
The SFRs of the both high-SFR and low-SFR groups of the bimodality seem to 
increase with M$_{\rm star}$. 
Since the bimodality is also seen in the case where the SFR$_{\rm corrected-UV}$ 
is used for all the sample, it is not due to the effect of the detection limit of the 
MIPS 24 $\mu$m data. 
While the median SFRs of galaxies at $2.5<z<3.5$ are similar with 
those at $1.5<z<2.5$ in all stellar mass bins, the average SFRs of these galaxies 
tend to be higher than those at $1.5<z<2.5$ due to galaxies with very high SFRs, 
especially at high mass.

%--------------------------figures
\begin{figure*}
\begin{center}
\includegraphics[angle=0,scale=1.0]{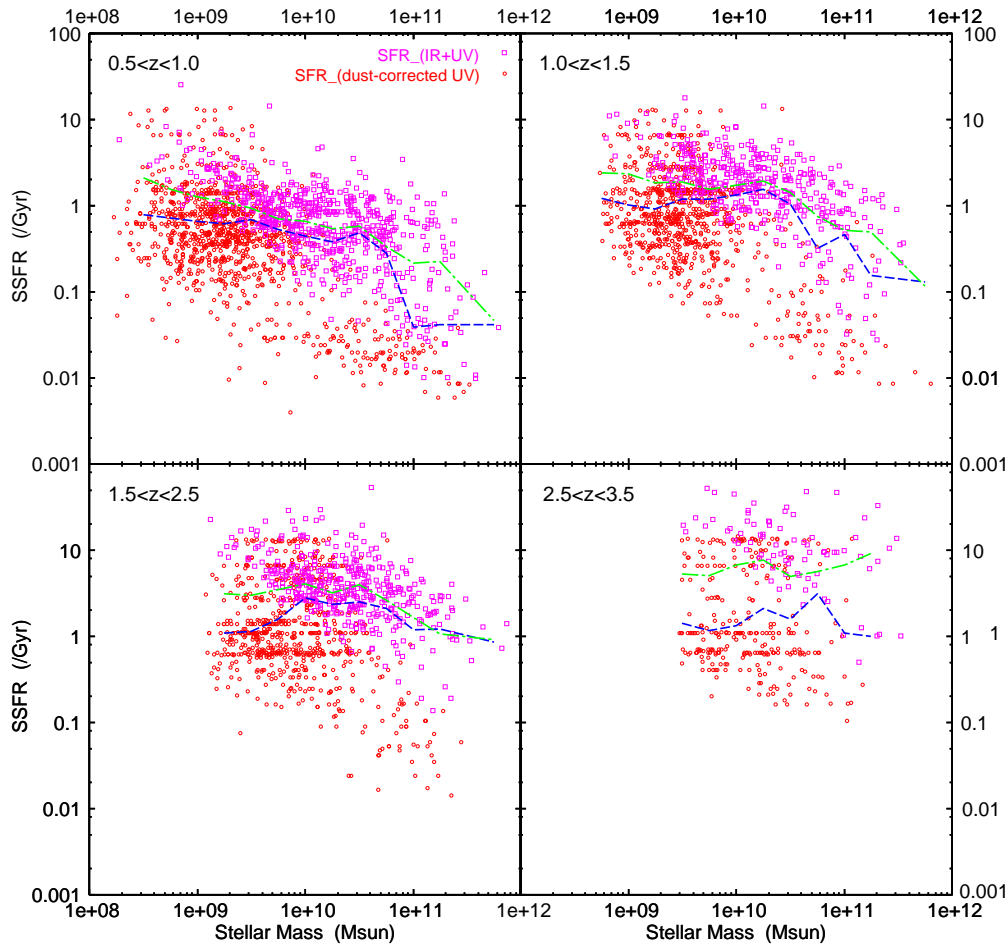}
\end{center}
\caption{Specific star formation rate (SFR/M$_{\rm star}$)  vs M$_{\rm star}$ 
for our stellar mass limited sample for 
each redshift bin. 
SFR$_{\rm IR+UV}$ is used if a galaxy is detected in the MIPS 24$\mu$m
 image and SFR$_{\rm corrected-UV}$ is used 
for objects without the 24$\mu$m detection.
Symbols are the same as Figure \ref{fig:sfrmsir}.
}
\label{fig:ssfrmsir}
\end{figure*}
%%%%%%%%%%%%%%%%%%%%%%%%%%%%%%%%%%%%
%--------------------------figures
\begin{figure*}
\begin{center}
\includegraphics[angle=0,scale=1.0]{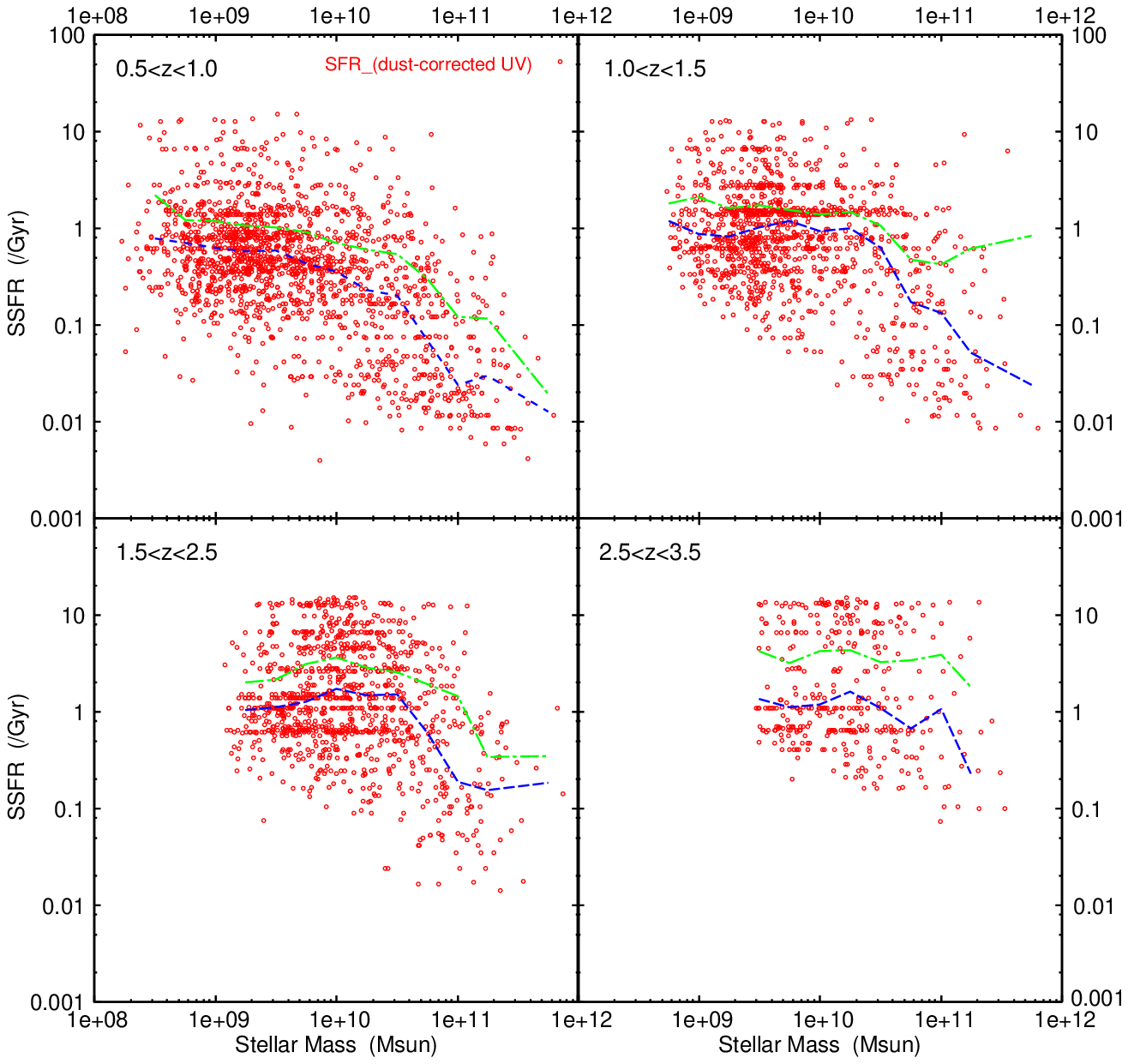}
\end{center}
\caption{The same as Figure \ref{fig:ssfrmsir} except that 
SFR$_{\rm corrected-UV}$ is used for all the sample.
}
\label{fig:ssfrmsuv}
\end{figure*}
%%%%%%%%%%%%%%%%%%%%%%%%%%%%%%%%%%%%
\subsection{SSFR  vs  M$_{\rm star}$}
\label{sec:ssfrms}
In Figure \ref{fig:ssfrmsir} and \ref{fig:ssfrmsuv}, 
we show SSFR vs M$_{\rm star}$ for our sample in the both cases with the 
different SFR indicators.
At M$_{\rm star} \lesssim 10^{10.5}$ M$_{\odot}$, 
the median and average of the SSFR are nearly independent of 
M$_{\rm star}$ at least at $z>1$, 
although the median and average values slightly decrease with 
M$_{\rm star}$ in the lowest redshift bin.  
This indicates that the relation between SFR and M$_{\rm star}$ for galaxies 
with M$_{\rm star} \lesssim 10^{10.5}$ M$_{\odot}$  has a logarithmic slope of 
$\sim$ 1 (slightly less than 1 for the lowest redshift bin). 
On the other hand, 
the median and average SSFRs decrease with M$_{\rm star}$ at 
M$_{\rm star} \gtrsim 10^{10.5}$ M$_{\odot}$, except for the average SSFR in 
the $2.5<z<3.5$ bin.

%--------------------------figures
\begin{figure}
\begin{center}
\includegraphics[angle=0,scale=0.8]{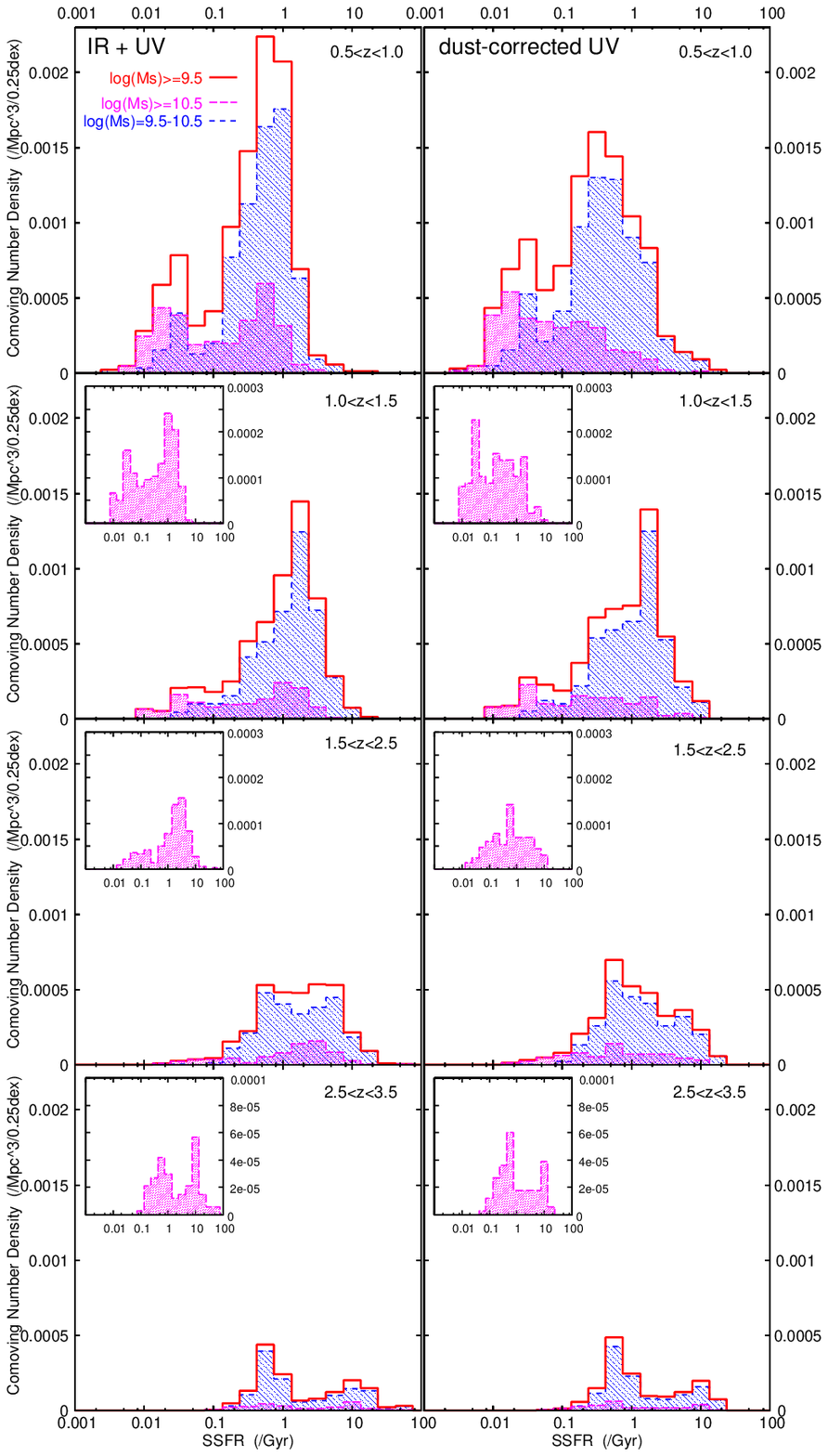}
\end{center}
\caption{Distribution of the specific star formation rate (SFR/M$_{\rm star}$)
for galaxies with M$_{\rm star} > 10^{9.5}$ M$_{\odot}$ in each redshift bin. 
Left panels show the case where SFR$_{\rm IR+UV}$ is used for galaxies detected 
at 24 $\mu$m and SFR$_{\rm corrected-UV}$ for the other objects, while 
right panels show the case where 
SFR$_{\rm corrected-UV}$ is used for all the sample. 
Long dashed line (Magenta) represents galaxies with M$_{\rm star} > 10^{10.5}$ 
M$_{\odot}$, and short dashed line (blue) shows those with M$_{\rm star} = 
10^{9.5-10.5}$ M$_{\odot}$. 
The insets represent the zoom-up of the plots for galaxies with M$_{\rm star} > 10^{10.5}$ 
M$_{\odot}$. 
}
\label{fig:ssfr}
\end{figure}
%%%%%%%%%%%%%%%%%%%%%%%%%
In order to investigate the evolution of the SSFR,  we show a distribution of SSFR 
 for the cases with the different SFR 
indicators in Figure \ref{fig:ssfr}.
We use the V$_{\rm max}$ method to calculate the comoving number density.
K09 calculated the V$_{\rm max}$ for the $K$-band magnitude limit 
($K=23$ for the wide field or $K=24$ for the deep field) 
by using  the best-fit model template for each object (see Section 3.3 in K09 
for details). 
In the figure, we use only galaxies with M$_{\rm star} > 10^{9.5}$ M$_{\odot}$ 
for a fair comparison among the different redshift bins. We also show  
galaxies with M$_{\rm star} = $ 10$^{9.5-10.5}$ M$_{\odot}$ and 
with M$_{\rm star} > 10^{10.5}$ M$_{\odot}$ separately.
It is seen that the distribution of the SSFR shifts to higher values with redshift, 
while the comoving number density of galaxies decreases. 

At $2.5<z<3.5$,  
the bimodality in the SFR mentioned above is also seen 
in the SSFR distribution, as a more simple form. 
There are a group of galaxies with high SSFRs of SFR/M$_{\rm star} \sim $ 10 
Gyr$^{-1}$ and that of galaxies with relatively low SSFRs of SFR/M$_{\rm star} \sim $ 
0.6 Gyr$^{-1}$. Thus,  a constant SSFR of SFR/M$_{\rm star} \sim $ 2 Gyr$^{-1}$ 
can divide between the two populations independent of M$_{\rm star}$ in the 
both cases with the different SFR indicators.

%The two highest mass bins
% at $1.0<z<1.5$ are affected strongly by one object with 
%SFR$_{\rm corrected-UV} > $ 1000 M$_{\odot}$/yr (Figure \ref{fig:sfrmsuv}) 
%and have large uncertainty. 

%--------------------------figures
\begin{figure}
\begin{center}
\includegraphics[angle=0,scale=0.65]{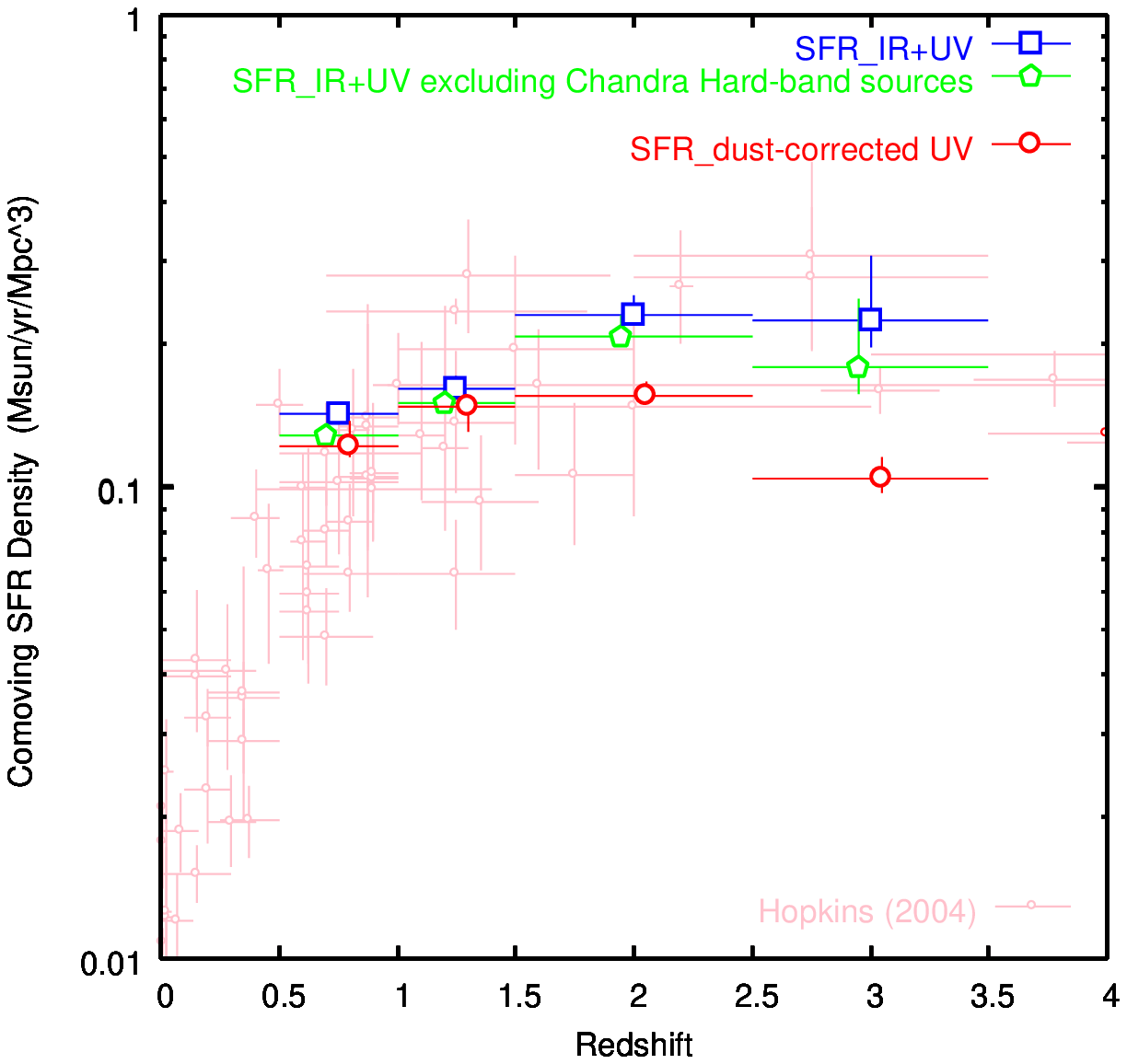}
\end{center}
\caption{Evolution of the total cosmic SFR density for 
our stellar mass limited sample for the cases with the different SFR 
indicators. The case with SFR$_{\rm IR+UV}$ 
where the Chandra hard-band sources are excluded 
is also shown (the result with SFR$_{\rm corrected-UV}$ does not change 
significantly if we exclude the Chandra hard-band sources).
Small open circles show the compilation from \cite{hop04}.
}
\label{fig:madau}
\end{figure}
%%%%%%%%%%%%%%%%%%%%%%%%%
\subsection{Contribution to the Cosmic SFR Density} 
\label{sec:sfrd}
Here we examine the contribution to the cosmic SFR density (hereafter, SFRD) 
of galaxies with different stellar masses.
We use the same V$_{\rm max}$ as used in Section \ref{sec:ssfrms} to 
estimate the SFRD. 
At first, we show the integrated SFRD for our sample and compare it with 
a compilation by \cite{hop04} in Figure \ref{fig:madau}.
The integrated SFRD was calculated from the sample limited by the stellar 
mass which depends on redshift (Section \ref{sec:sample}) and was not 
extrapolated down to lower mass. Therefore the SFRD at higher redshift in 
the figure was integrated only down to the higher stellar mass limit.
Since the contribution of galaxies with $M_{\rm star} < 10^{10}$ M$_{\odot}$ 
seems to be negligible even at high redshift as we show below, however, 
the effect of the systematically higher mass limit at high redshift 
is expected to be small. 

In Figure \ref{fig:madau}, we plot the both results with the different 
SFR indicators and also show the case with SFR$_{\rm IR+UV}$ 
where Chandra hard-band sources are excluded. 
The integrated SFRDs for our sample increase with redshift up to $z\sim 2$ 
and are constant or slightly decrease between $z\sim 2$ and $z \sim 3$.
 They range within the compilation by 
\cite{hop04} over $0.5<z<3.5$, except for a lower SFRD  
at $z>1.5$ in the case with SFR$_{\rm corrected-UV}$ 
for all the sample. 
$SFR_{\rm corrected-UV}$ for galaxies at high redshift might be 
underestimated due to the heavy obscuration (Section \ref{sec:compsfr}). 

%--------------------------figures
\begin{figure*}
\begin{center}
\includegraphics[angle=0,scale=1.15]{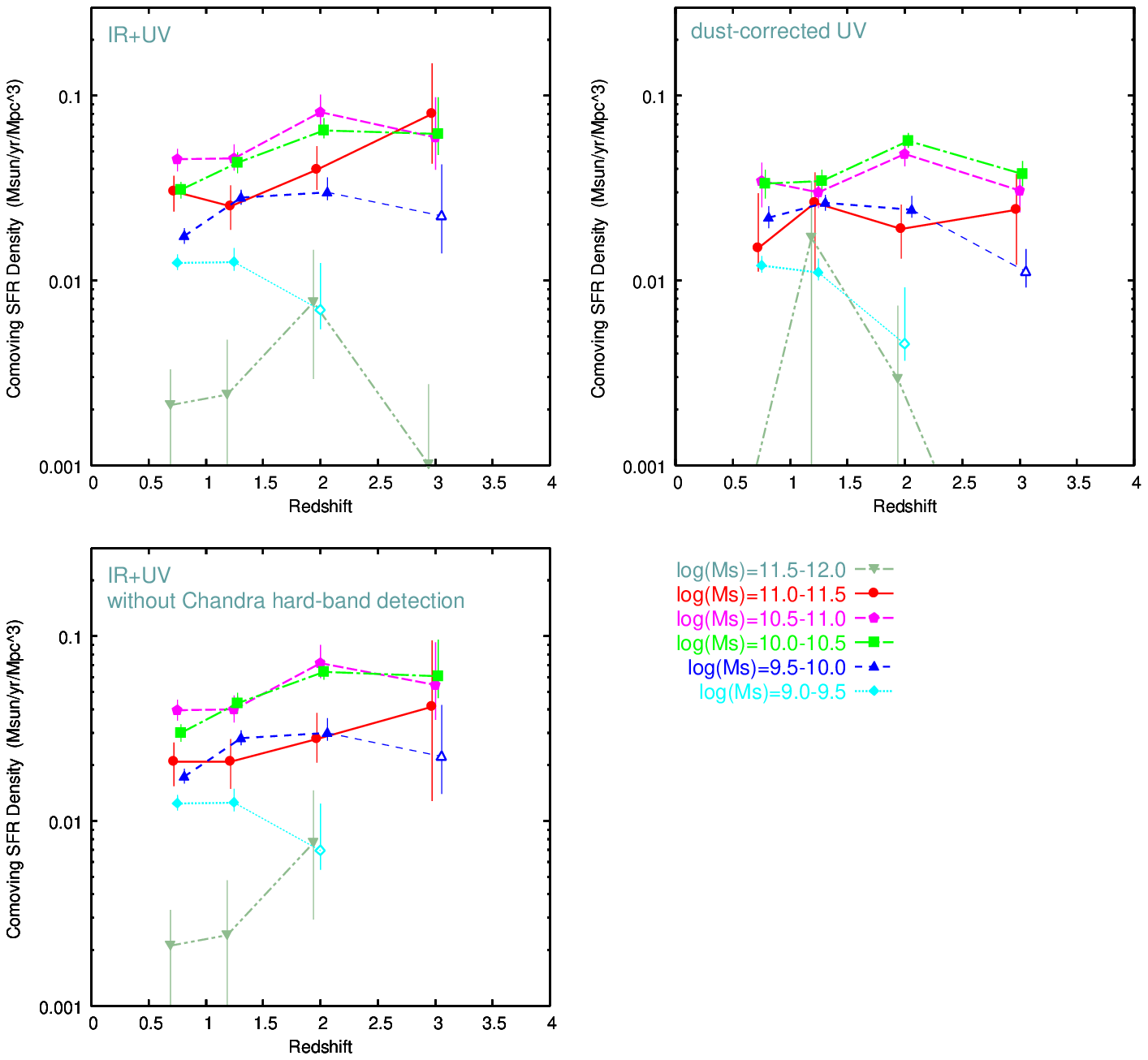}
\end{center}
\caption{Contributions of galaxies in different stellar mass ranges to 
the cosmic SFR density as a function of redshift.
Upper left and right panels show the cases with the different SFR indicators. 
Bottom left panel represents the case with SFR$_{\rm IR+UV}$ 
where the Chandra hard-band sources are excluded. 
Open symbols mean that the incompleteness due to the stellar mass limit could be 
significant for the redshift bin.
}
\label{fig:sfrdms}
\end{figure*}
%%%%%%%%%%%%%%%%%%%%%%%%%
Figure \ref{fig:sfrdms} shows the contribution to the cosmic SFRD of 
galaxies in the different ranges of stellar mass. 
In the all redshift range, galaxies with M$_{\rm star} = 10^{10-11}$ M$_{\odot}$ 
dominate the cosmic SFRD; their contribution is about 1/2 -- 2/3 of the total
 SFRD. 
 The increase of the cosmic 
SFRD with redshift up to $z\sim2$ seems to reflect
 mainly the evolution of the contribution of galaxies in this mass range.

In the case with SFR$_{\rm IR+UV}$, the contribution of 
massive galaxies with M$_{\rm star} = 
10^{11-11.5}$ M$_{\odot}$ shows stronger evolution at $z\gtrsim 1.5$ than 
that of galaxies with M$_{\rm star} = 10^{10-11}$ M$_{\odot}$ and 
continues to increase up to $z\sim3$. 
These massive galaxies significantly contribute the cosmic SFRD at $2.5<z<3.5$, 
while the contribution of these galaxies is relatively small at $z\lesssim 2$.
In the case where the Chandra hard-band sources are excluded, 
the contribution of these massive galaxies is slightly lower and its evolution 
is milder, although the contribution becomes significant at $z\sim3$ 
even in this case. This is because the X-ray AGN is preferentially 
associated with 
massive galaxies (e.g., \citealp{bes05}) and 
because the fraction of objects detected at hard 
X-ray in massive galaxies becomes very high at $z\gtrsim 2$ (\citealp{yam09}; 
\citealp{bru09}).
While the SFR$_{\rm IR+UV}$ of the Chandra hard-band sources could be 
overestimated due to the contribution of AGN, the exclusion of these galaxies 
might cause the underestimate of the contribution of massive galaxies 
to the SFRD if a significant fraction of the IR luminosity of these galaxies 
arises from star formation activities.
In the case where SFR$_{\rm corrected-UV}$ is used for all the sample, 
the contribution of these galaxies with M$_{\rm star} = 10^{11-11.5}$ M$_{\odot}$
has a large uncertainty even at low redshift and its evolution  
is less clear. This is partly because the large error of the SFR 
for heavily obscured galaxies with high SFR. In particular, the contribution at 
$1.0<z<1.5$ is strongly affected one such object with 
SFR$_{\rm corrected-UV} > $ 1000 M$_{\odot}$/yr (Figure \ref{fig:sfrmsuv}). 

The contribution of galaxies with M$_{\rm star} < 10^{10}$ M$_{\odot}$ seems 
to be small at any redshift, although the completeness limit for 
our sample becomes $\sim 10^{9.5-10}$ M$_{\odot}$ at $z\sim 3$.
The contribution of galaxies with M$_{\rm star} = 10^{9.5-10}$ M$_{\odot}$ 
is $\sim$ 15\% of the total SFRD at $0.5<z<3.5$ and evolves
 similarly with that for galaxies with M$_{\rm star} = 10^{10-11}$ M$_{\odot}$. 
Galaxies with M$_{\rm star} > 10^{11.5}$ M$_{\odot}$ have a negligible contribution, 
although there are only a few such very massive galaxies in each redshift bin 
and the uncertainty of the contribution is very large.

\section{Discussion}
\label{sec:discuss}

\subsection{Comparison with Other Studies at $z<2.5$}
\label{sec:compdis}
We have investigated the distribution of SFR as a function of stellar mass 
for galaxies at $0.5<z<3.5$, using the stellar mass limited sample 
based on the very deep NIR imaging data of the MODS.
We here compare our results in the MODS field with previous studies mainly 
at $z<2.5$.

Several studies examined the relation between SFR and stellar mass of star-forming 
galaxies, and found that more massive galaxies tend to have higher SFRs 
at $z\sim1$ (\citealp{noe07a}; \citealp{elb07}) and at $z\sim2$ (\citealp{san09}; 
\citealp{dad07}; \citealp{pan09}; \citealp{dun09}). 
At M$_{\rm star} \lesssim 10^{10.5}$ M$_{\odot}$, we found a similar trend in our 
 sample at $0.5<z<2.5$, and confirmed the same trend 
at $2.5<z<3.5$ although the SFR distribution 
shows the bimodality and does 
not concentrate around the average value.
At M$_{\rm star} \gtrsim 10^{10.5}$ M$_{\odot}$, there are massive galaxies with 
similar SFRs with lower-mass galaxies in our stellar mass-selected sample, 
while galaxies with higher SFRs which follow the trend at lower M$_{\rm star}$ also 
exist. If we excluded galaxies with relatively low SFRs as a quiescent population, 
we could consider the relation between SFR and stellar mass continues to high 
mass for star-forming galaxies, which has been reported in other studies. 

Our result that the SFR (SSFR) distribution at a fixed M$_{\rm star}$ 
shifts to higher values with increasing redshift
is also consistent with  these previous studies. 
For example, 
\cite{pan09} estimated an average SFR as a function of M$_{\rm star}$ for 
star-forming BzK galaxies with $K\lesssim$ 21.2 
at $z\sim2$ in the COSMOS field 
by performing a stacking analysis of 1.4 GHz radio continuum. 
They pointed out that the SSFR of these star-forming galaxies show a evolution of 
a factor of $\sim$ 4 between $z=1.4$ and $z=2.3$. 
Using the stellar mass-selected sample 
 whose limiting mass reaches $\sim$ 10$^{9.5}$--10$^{10}$ 
 M$_{\odot}$ even at $z\sim3$, we found that 
more massive galaxies show stronger evolution of the SFR at $z\gtrsim 1$. 

We found the median and average SSFRs of galaxies is nearly independent of 
M$_{\rm star}$ at M$_{\rm star} \lesssim 10^{10.5}$ M$_{\odot}$ at $z>1$,  
while they slightly decrease with M$_{\rm star}$ at $0.5<z<1.0$. 
This indicates that 
the logarithmic slope of the relation between SFR and M$_{\rm star}$ 
is $n \sim 1$ ($n \lesssim 1$ for $0.5<z<1.0$). 
Other studies reported the similar slope of the SFR-M$_{\rm star}$ relation, 
although there is some variance among the studies in the observed slope and 
scatter of the relation (\citealp{dad07}; \citealp{san09}; \citealp{pan09}; 
\citealp{elb07}; \citealp{noe07a}).
The variance might be explained by differences in the sample selection or 
the SFR indicators among the studies. 

The SSFR independent of M$_{\rm star}$ at M$_{\rm star} < 10^{10.5}$ 
M$_{\odot}$ 
means that the stellar mass growth by star formation activities in
each galaxy is similar among galaxies with different masses.
Using the same sample as this study, K09 investigated the evolution of the
galaxy stellar mass function (SMF), and found that the low-mass slope of the
SMF becomes steeper with redshift at $1 \lesssim z \lesssim 3$.
The nearly constant mass growth rate by
star formation expected from the slope of n $\sim$ 1 implies that star formation
in each galaxy does not significantly change
the low-mass slope with time.
If such mass-independent SSFR continues down to lower mass than the stellar
mass limit of our sample in the high redshift bins,
additional mechanisms such as the hierarchical merging, which destroys small
galaxies and builds more massive ones, might be needed in order to
reproduce the evolution of the low-mass slope (K09).

%In summary, our results of the distribution of the SFR as a function of 
%stellar mass and its evolution are broadly consistent with previous studies.
%While the normalization of the SFR-M$_{\rm star}$ relation generally 
%agrees well at each redshift, 
%the slope of the relation seems to show some variance 
%among the studies, which might be due to the selection effect and/or  
%the systematic differences in the SFR indicators.
 
%--------------------------figures
\begin{figure*}
\begin{center}
\includegraphics[angle=0,scale=0.9]{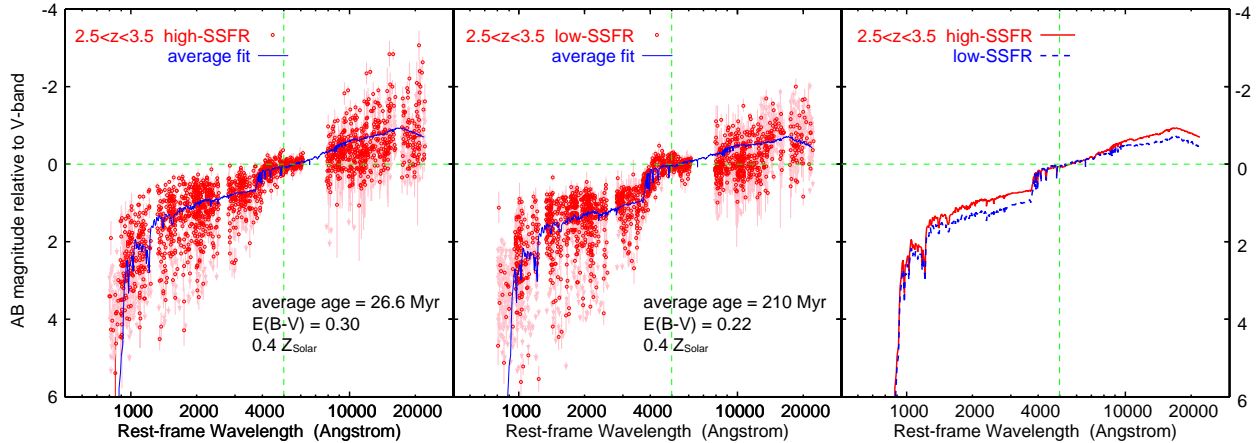}
\end{center}
\caption{Rest-frame broad-band SEDs of star-forming galaxies at $2.5<z<3.5$.
Left and middle panels show those with 
$\log$(SFR$_{\rm IR+UV}$/M$_{\rm star}$) $>$ 0.25 
(the high-SSFR population) and
 those with $\log$(SFR$_{\rm IR+UV}$/M$_{\rm star}$) $=$ -0.5 -- 0.25 
(the low-SSFR population).
Errorbars represent the 1$\sigma$ photometric uncertainties and arrows indicate 
the 2$\sigma$ upper limits.
Solid line in each panel  shows the best-fit SED model for all the data points. 
Average stellar age, color excess, and metallicity of the best-fit model are also shown 
in the panels.
Right panel shows a comparison between these average SEDs for the high- and 
low-SSFR populations. 
If we use the SFR$_{\rm corrected-UV}$ for all the sample, 
the best-fit SEDs for the high- and low-SSFR populations
are almost the same as seen in this figure. 
}
\label{fig:sedssfr}
\end{figure*}
%%%%%%%%%%%%%%%%%%%%%%%%%
\subsection{Bimodality in SSFR of star-forming galaxies at $2.5<z<3.5$}
\label{sec:sfrmsdis}
We found the bimodality in the SSFR distribution of galaxies at $2.5<z<3.5$, which
can be divided into two populations by a constant SSFR of $\sim$ 2 Gyr$^{-1}$. 
One population consists of galaxies with SFR/M$_{\rm star} \sim $ 0.5--1 Gyr$^{-1}$, 
and those with SFR/M$_{\rm star} \sim $ 10 Gyr$^{-1}$ belong to the other 
population. 
For the SSFR of galaxies at $z>2.5$,  
the SFR and M$_{\rm star}$ of Lyman Break Galaxies (LBGs) have been investigated 
by several studies. 
\cite{mag10} reported that LBGs  detected in the 
IRAC 3.6 $\mu$m and 4.5 $\mu$m bands at $z\sim3$ show the SFR-M$_{\rm star}$ 
relation with a slope of n $\sim$ 0.9 and a considerable scatter, 
and that their median SSFR of $\sim$ 
4.5 Gyr$^{-1}$ is higher than those at lower redshifts, 
while several observational studies suggest that 
the SFR-M$_{\rm star}$ relation for Lyman Break Galaxies 
at $4 \lesssim z \lesssim 7$ is similar with that at $z\sim2$ 
(\citealp{dad09}; \citealp{sta09}; \citealp{gon10}).
The range of the SSFR of LBGs in \cite{mag10} is similar with galaxies at 
$2.5<z<3.5$ in our sample, although most of LBGs in their sample have  
M$_{\rm star} \gtrsim 10^{10.5}$ M$_{\odot}$ probably due to the IRAC selection and 
the number of galaxies with low SSFR is relatively small. 

\cite{dut09} constructed a semi-analytic model for disk galaxies 
and suggested that 
the evolution of the normalization of the SFR-M$_{\rm star}$ 
relation for these disk galaxies at $z \lesssim 2$ 
closely follows the evolution in the gas (and dark matter) accretion rate.  
They predicted that the normalization continues to increase 
with redshift up to $z>2$, if the SFR follows the gas accretion rate at 
such high redshift. 
The SFR of the high-SSFR population at $2.5<z<3.5$ 
may be related with the evolution of the gas accretion rate at 
such high redshift, although the SSFR of $\sim$ 10 Gyr$^{-1}$ at $z\sim3$ is 
higher than the model prediction by \cite{dut09}.

Recently, \cite{dad10} reported a bimodality in 
 the SFR relative to the gas mass 
for star-forming galaxies at low and high redshifts and proposed that 
the star formation efficiency would be 
 different between disk and starburst galaxies 
because different modes of star formation work in the two populations. 
The high-SSFR population 
at $2.5<z<3.5$ in our sample might correspond to 
such starburst galaxies. 
%%%%%%%%%%%%%%%%%%%%%%%%%%%%%%%%%%
For example, the starburst sequence consists of sub-millimeter galaxies (SMGs) 
at high redshift  
 as well as local ULIRGs in \cite{dad10} and several studies suggest that 
high-z SMGs tend to show relatively 
high SSFRs (e.g., \citealp{tak08}; \citealp{dad09}).
There are 23 SCUBA sources in the MODS field \citep{pop06}, and 
all but one (GN10, \citealp{wan07}; \citealp{wan09}) have counterparts in 
the $K$-selected catalog of the MODS. 
Of 22 sources detected in the $K$-band image, 
 6 SCUBA sources lie at $2.5<z<3.5$ (3 with spectroscopic 
redshift and 3 only with photometric redshift), 
all but one hard X-ray source (GN22, \citealp{lai10}) belong to  
the high-SSFR population with SFR/M$_{\rm star} \sim$ 10 Gyr$^{-1}$. 
Figure \ref{fig:sedssfr} shows the rest-frame broad-band SEDs of star-forming galaxies 
at $2.5<z<3.5$ on the high- and low-SSFR populations. 
We defined the high-SSFR population as 
galaxies with  $\log$(SFR/M$_{\rm star}$[Gyr$^{-1}$]) $>$ 0.25 and the low-SSFR 
population as those with  $\log$(SFR/M$_{\rm star}$) $=$ -0.5--0.25, respectively.
The high-SSFR population shows weaker Balmer break feature 
and redder color 
at long wavelength, which indicates that these galaxies are younger and dustier, 
while the overall slope of SEDs and scatters among the objects are similar 
between the two populations. 
Those objects in the high-SSFR population might 
have higher star formation efficiency than ordinary star-forming 
(disk) galaxies.  
%%%%%%%%%%%%%%%%%%%%%%%%%%%%%%%%%%

\subsection{Evolution of Cosmic SFRD and Stellar Mass Assembly of Galaxies}
\label{sec:sfrddis}
In Section \ref{sec:sfrd}, we saw that galaxies with 
M$_{\rm star} = 10^{10-11}$ M$_{\odot}$ 
dominate the cosmic SFRD at $0.5<z<3.5$, and the contribution of 
galaxies with  M$_{\rm star} = 10^{11-11.5}$ M$_{\odot}$ increases 
with redshift even at $z\gtrsim2$ and becomes significant at $z\sim3$.
On the other hand, K09 investigated the contribution of galaxies in 
different mass ranges to the cosmic stellar mass density, and found that 
galaxies with M$_{\rm star} \sim 10^{10.5-11.5}$ M$_{\odot}$ 
($10^{10-11.5}$ M$_{\odot}$ at $z\sim3$) dominate the cosmic stellar mass 
density (Figure 17 in K09).
Considering the nearly constant SSFR among (star-forming) 
galaxies with different stellar masses discussed in Section \ref{sec:compdis}, 
the result that galaxies with $10^{10-11.5}$ M$_{\odot}$ dominate the 
cosmic SFRD is qualitatively consistent with 
the dominant contribution of these galaxies in the stellar mass density. 
Major star formation in the universe at $1 \lesssim z \lesssim 3$ 
seems to occur in galaxies with stellar mass 
which dominates the cosmic stellar mass density 
or those with slightly lower stellar mass. 

%--------------------------figures
\begin{figure}
\begin{center}
\includegraphics[angle=0,scale=0.65]{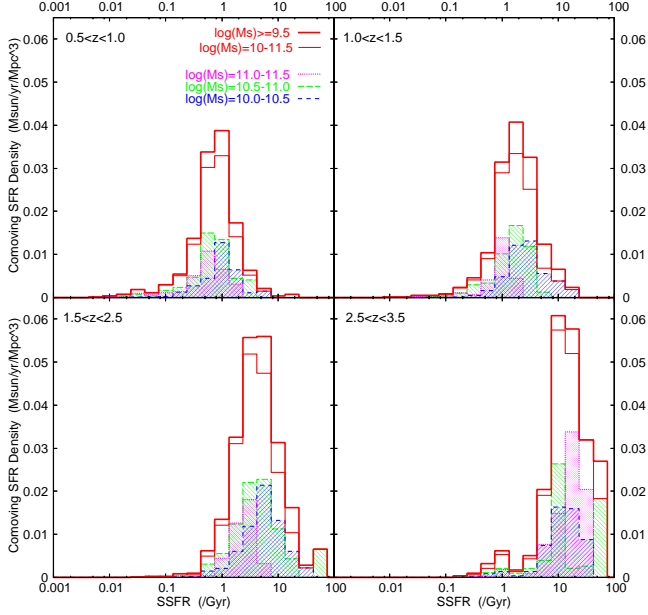}
\end{center}
\caption{Contributions of galaxies in different SSFR ranges to the cosmic 
SFR density in each redshift bin for the case 
where SFR$_{\rm IR+UV}$ is used for galaxies detected 
at 24 $\mu$m and SFR$_{\rm corrected-UV}$ for the other objects.  
Only galaxies with M$_{\rm star} > 10^{9.5}$ M$_{\odot}$ are used for all the 
redshift bins. Thick solid line shows galaxies with M$_{\rm star} > 10^{9.5}$ M$_{\odot}$ 
and thin solid line represents galaxies with M$_{\rm star} = 10^{10-11.5}$ M$_{\odot}$. 
Dotted (magenta), long dashed (green), and short dashed (blue)
lines represent galaxies with M$_{\rm star} = 10^{11-11.5}$, $10^{10.5-11}$, and 
$10^{10-10.5}$ M$_{\odot}$, respectively. 
}
\label{fig:sfrdssfrir}
\end{figure}
%%%%%%%%%%%%%%%%%%%%%%%%%
The cosmic SFRD increases with redshift from $z\sim0$ up to $z\sim2$ and 
is roughly constant or slightly decreases between $z\sim2$ and $z\sim3$ 
(e.g., \citealp{hop06}).
The evolution of the cosmic SFRD roughly follows that of galaxies
with M$_{\rm star} = 10^{10-11}$ M$_{\odot}$, which dominate the SFRD  
(Figure \ref{fig:madau} and \ref{fig:sfrdms}).
In Figure \ref{fig:histsfrir} and \ref{fig:histsfruv}, 
the average SFR per galaxy for those with 
M$_{\rm star} = 10^{10-11}$ M$_{\odot}$ increases with redshift up to $z\sim 3$, 
but the evolution between $z\sim 2$ and $z\sim3$ seems to be weaker 
than that at lower redshift.
On the other hand, K09 investigated the evolution of the number density of 
galaxies in different mass ranges, and showed that the number density 
of these galaxies decreases gradually with redshift (Figure 16 in K09).
At $z\lesssim 2$, the increase in the average SFR per galaxy with redshift 
overwhelms the decrease in the number density, and therefore the contribution 
of these galaxies to the total SFRD increases with redshift. 
At $2 \lesssim z \lesssim 3$, the evolution of the SFR per galaxy becomes 
weaker, while the number density continues to decrease even at such high 
redshift. Then these two roughly balances,
 and the SFRD is nearly constant or slightly decreases with redshift.
For galaxies with M$_{\rm star} = 10^{11-11.5}$ M$_{\odot}$,  
the average SFR evolves more strongly between $z\sim2 $ and $z\sim3$ 
especially in the case with SFR$_{\rm IR+UV}$.
The strong evolution of the SFR per galaxy for these massive galaxies continues 
to overwhelm the decrease in the number density up to $z\sim3$, 
although the number density of these galaxies also continues to decrease
to $z\sim3$. As a result, the contribution of these galaxies with 
M$_{\rm star} = 10^{11-11.5}$ M$_{\odot}$ to the SFRD increases with redshift
even at $z\gtrsim 2$. 
%--------------------------figures
\begin{figure}
\begin{center}
\includegraphics[angle=0,scale=0.65]{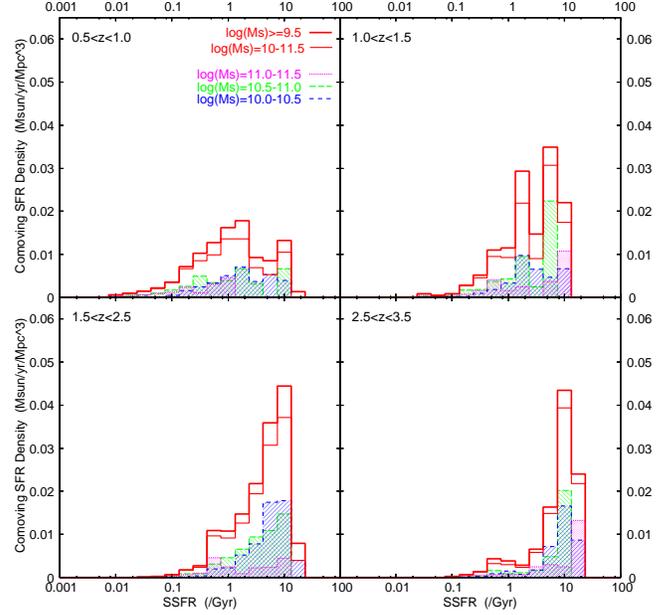}
\end{center}
\caption{The same as Figure \ref{fig:sfrdssfrir} but for the case where
SFR$_{\rm corrected-UV}$ is used for all the sample.
}
\label{fig:sfrdssfruv}
\end{figure}
%%%%%%%%%%%%%%%%%%%%%%%%%

In other words, the number density of galaxies is relatively small, but 
the average SFR per galaxy is high at $z\sim3$.
As time goes on, the SFR in each galaxy decreases, while the number density 
of galaxies at a fixed M$_{\rm star}$  increases due to the stellar mass 
growth in galaxies with a wide range of masses. 
In particular, the SFR in massive galaxies with  
M$_{\rm star} = 10^{11-11.5}$ M$_{\odot}$ decreases rapidly, 
and  a fraction of these galaxies enters into a quiescent phase (massive galaxies 
with low SSFR in Figure \ref{fig:ssfrmsir} and \ref{fig:ssfrmsuv}). 
 As a result, the contribution of these galaxies to the SFRD 
becomes relatively small by $z\sim2$. 
At $z \lesssim 2$, the decrease of the SFR in each galaxy with 
 M$_{\rm star} = 10^{10-11}$ M$_{\odot}$ overwhelms the increase in the 
number density, which drives the overall decrease of the cosmic SFRD with time.

In Figure \ref{fig:sfrdssfrir} and \ref{fig:sfrdssfruv}, 
we show the contribution of galaxies with different SSFRs to the cosmic 
SFRD in the cases with SFR$_{\rm IR+UV}$ and SFR$_{\rm corrected-UV}$, respectively. 
It is seen that in galaxies with M$_{\rm star} = 10^{10-11.5}$ M$_{\odot}$, 
those with a relatively narrow range of SSFR 
($\lesssim 1$ dex) dominate the cosmic SFRD, especially in the case with 
SFR$_{\rm IR+UV}$. 
Furthermore, the SSFR of galaxies which dominate the SFRD systematically 
increases with redshift. 
In this context, major star formation in the universe at higher redshift seems to be 
associated with a more rapid growth of stellar mass of galaxies. 
At $2.5<z<3.5$,  in particular, 
%galaxies with M$_{\rm star} = 10^{10-11.5}$ M$_{\odot}$, 
the high-SSFR population with 
SFR/M$_{\rm star} \sim $ 10 Gyr$^{-1}$, which are relatively small in 
number (Figure \ref{fig:ssfr}), dominate the SFRD. 
If these objects would be the 'starburst' galaxies with distinctively higher 
star formation efficiency than ordinary disk galaxies  
as discussed in the previous section, 
these galaxies are expected to grow their stellar mass rapidly and could cause 
the increase of the number density of massive galaxies at $1 \lesssim z \lesssim 3$. 
There might be a transition in the dominant component of the 
 star formation in the universe 
between $z\sim3$ and $z\sim1$; from starburst in relatively small number 
of galaxies at $z\sim3$, which drives a rapid 
growth of stellar mass, to 
 star formation in many disk galaxies at $z\sim1$, 
which leads to a more gradual growth of stellar mass.

%% The displaymath environment will produce the same sort of equation as
%% the equation environment, except that the equation will not be numbered
%% by LaTeX.

\section{Summary}
We studied SFR as a function of M$_{\rm star}$ for galaxies at 
$0.5<z<3.5$ in the GOODS-North field, using the $K$-selected sample  
from the MODS. 
The very deep NIR data of the MODS allow us to construct the stellar mass 
limited sample down to 10$^{9.5-10}$ M$_{\odot}$ even at $z\sim3$.
The rest-frame 2800 \AA\ luminosity and the MIPS 
24 $\mu$m flux were used to estimate the SFRs of the sample galaxies. 
Using these SFR indicators, we showed the two cases of the results, namely, 
one case where SFR$_{\rm IR+UV}$ is used if a galaxy is detected in the MIPS 24$\mu$m
 image and SFR$_{\rm corrected-UV}$ is used 
for those without the 24$\mu$m detection, and   
the other case where SFR$_{\rm corrected-UV}$ is used for all the sample. 

Main results in this study are as follows.
\begin{itemize}

\item More massive galaxies tend to have higher SFRs at M$_{\rm star} \lesssim
 10^{10.5}$ M$_{\odot}$ in all redshift ranges. 
At M$_{\rm star} \gtrsim 10^{10.5}$ M$_{\odot}$, 
there are galaxies with relatively low SFRs similar with low-mass galaxies, 
while galaxies with higher SFRs which follow the trend at lower M$_{\rm star}$ also 
exist. 

\item The median and average SSFRs of galaxies is nearly independent of 
M$_{\rm star}$ at M$_{\rm star} \lesssim 10^{10.5}$ M$_{\odot}$ at $z>1$,  
while they slightly decrease with M$_{\rm star}$ at $0.5<z<1.0$. 
At M$_{\rm star} \gtrsim 10^{10.5}$ M$_{\odot}$, 
the median and average SSFRs decrease with M$_{\rm star}$. 

\item The SFR (SSFR) distribution at a fixed M$_{\rm star}$ 
shifts to higher values with increasing redshift. 
More massive galaxies show the stronger evolution of SFR at $z\gtrsim 1$. 

\item At $2.5<z<3.5$, the bimodality in the SFR and SSFR distributions 
at each mass is seen. 
The bimodality is the more simple form in SSFR and   
can be divided into the two populations by a constant SSFR of $\sim$ 2 Gyr$^{-1}$. 
One population consists of galaxies with SFR/M$_{\rm star} \sim $ 0.5--1 Gyr$^{-1}$, 
and the other population is the high-SSFR population, to which 
galaxies with SFR/M$_{\rm star} \sim $ 10 Gyr$^{-1}$ belong. 
These high- and low-SSFR
 populations might have different modes of star formation, namely,  
starburst and more continuous star formation in disks. 

\item Galaxies with M$_{\rm star} = 10^{10-11}$ M$_{\odot}$ dominate the 
cosmic SFRD at $0.5<z<3.5$. The evolution of these galaxies seems to 
drive the evolution of the cosmic SFRD, especially at $z \lesssim 2.5$.
The contribution of galaxies
with M$_{\rm star} = 10^{11-11.5}$ M$_{\odot}$ increases strongly 
with redshift at $z>1.5$ and becomes significant at $z\sim3$.
The contribution of galaxies with M$_{\rm star} < 10^{10}$ M$_{\odot}$ is 
relatively small at any redshift.

\item In galaxies with M$_{\rm star} = 10^{10-11.5}$ M$_{\rm star}$, 
those with a relatively narrow 
range of SSFR ($\lesssim$ 1 dex) dominate the cosmic SFR
at $0.5<z<3.5$ especially in the case 
with SFR$_{\rm IR+UV}$.
The SSFR of galaxies which dominate the SFRD 
systematically increases with redshift. 
At $2.5<z<3.5$, the high-SSFR population dominates the cosmic SFRD.  
Major star formation in the universe at higher redshift seems to be 
associated with a more rapid growth of stellar mass of galaxies.

\end{itemize}

Systematic errors in the SFRs estimated from the 24 $\mu$m flux and 
the optical-NIR broad-band SED (Section \ref{sec:compsfr}) 
could affect some of the above results. 
More precise measurements of SFR are essential  
to conclusively confirm the results in this study 
and to reveal the stellar mass assembly 
histories of galaxies at $1\lesssim z \lesssim 3$. For example, 
NIR spectroscopic surveys with sufficiently 
large sample for the stacking analysis and/or long exposure time  
enable to estimate directly the dust extinction for the nebular emission 
lines from the Balmer decrement, 
and to determine the dust-corrected H$\alpha$ luminosity 
of star-forming galaxies at that epoch with high accuracy. 
The forthcoming Herschel and ALMA 
observations will also provide strong constraints on the shape of 
the far-IR SED of these galaxies, which allows us to 
determine the total IR luminosity more precisely.


%% Included in this acknowledgments section are examples of the
%% AASTeX hypertext markup commands. Use \url without the optional [HREF]
%% argument when you want to print the url directly in the text. Otherwise,
%% use either \url or \anchor, with the HREF as the first argument and the
%% text to be printed in the second.

\acknowledgments

We thank an anonymous referee for very helpful suggestions and comments.
This study is based on data collected at Subaru Telescope, which is operated by
the National Astronomical Observatory of Japan. 
This work is based in part on observations made with the Spitzer Space
Telescope, which is operated by the Jet Propulsion Laboratory,
California Institute of Technology under a contract with NASA.
Some of the data presented in this paper were obtained from the Multi-mission
Archive at the Space Telescope Science Institute (MAST).
STScI is operated by the Association of Universities for Research in
Astronomy, Inc., under NASA contract NAS5-26555.
Support for MAST for non-HST data is provided by the NASA Office of
Space Science via grant NAG5-7584 and by other grants and contracts.
Data  reduction and analysis 
were carried out on common use data analysis computer system 
 at the Astronomy Data Center, ADC, of the National Astronomical 
Observatory of Japan.
IRAF is distributed by the National Optical Astronomy Observatories,
which are operated by the Association of Universities for Research
in Astronomy, Inc., under cooperative agreement with the National
Science Foundation.

\end{document}